\documentclass[smallcondensed]{svjour3}     
\smartqed  
\usepackage{graphicx}
\usepackage[authoryear]{natbib}
\usepackage{amsmath}
\newcommand{\echo}{EChO}
\newcommand{\echosim}{EChOSim}
\journalname{Experimental Astronomy}
\begin{document}

\title{Generation of an optimal target list for the Exoplanet Characterisation Observatory (EChO)}

\titlerunning{Generation of an optimal target list for EChO}        

\author{R. Varley \and
I. Waldmann \and
E. Pascale \and 
M. Tessenyi \and
M. Hollis \and 
J. C. Morales \and 
G. Tinetti \and
B. Swinyard \and
P. Deroo \and
M. Ollivier \and
G. Micela
}


\institute{R. Varley \and I. Waldmann \and M. Tessenyi \and M. Hollis \and G. Tinetti \and B. Swinyard \at
              University College London, Gower Street, WC1E 6BT, UK\\
              \email{r.varley@ucl.ac.uk}           
           \and
           E. Pascale \at
              School of Physics Astronomy, Cardiff University, Cardiff, CF24 3AA, UK
           \and
           J. C. Morales \at
              LESIA-Observatoire de Paris, CNRS, UPMC Univ. Paris 06, Univ. Paris-Diderot, 5 Pl. Jules Janssen, 92195 Meudon CEDEX, France
           \and
           P. Deroo \at
              Jet Propulsion Laboratory, California Institute of Technology, 4800 Oak Grove Drive, Pasadena, California 91109-8099 USA.
           \and
           M. Ollivier \at
           Institut d'Astrophysique Spatiale, Batiment 121, UniversiteÌ de Paris-Sud, 91405 ORSAY Cedex, France
           \and
           G. Micela \at
              INAF-Osservatorio Astronomico di Palermo, Italy
}

\date{Received: date / Accepted: date}

\maketitle

\begin{abstract}
The Exoplanet Characterisation Observatory (\echo) has been studied as a space mission concept by the European Space Agency in the context of the M3 selection process.
 Through direct measurement of the  atmospheric chemical composition of hundreds of exoplanets, EChO would address fundamental questions such as: What are exoplanets made of? How do planets form and evolve? What is the origin of exoplanet diversity?

More specifically, EChO is a dedicated survey mission for transit and eclipse spectroscopy capable of observing a large, diverse and well-defined planetary sample within its four to six year mission lifetime.

In this paper we use the end-to-end instrument simulator \echosim{ }to model the currently discovered targets, to gauge which targets are observable and assess the \echo{ }performances obtainable for each observing tier and time. We show that EChO would be capable of observing over 170 relativity diverse planets if it were launched today, and the wealth of optimal targets for EChO expected to be discovered in the next 10 years by space and ground-based facilities is simply overwhelming.

In addition, we build on previous molecular detectability studies to show what molecules and abundances will be detectable by \echo{ }for a selection of real targets with various molecular compositions and abundances.

\echo's unique contribution to exoplanetary science will be in identifying the main constituents of hundreds of exoplanets in various mass/temperature regimes, meaning that we will be looking no longer at individual cases but at populations. Such a universal view is critical if we truly want to understand the processes of planet formation and evolution in various environments.

In this paper we present a selection of key results. The full results are available online ({http://www.ucl.ac.uk/exoplanets/echotargetlist/}).
\end{abstract}

\section{Introduction}
Within the last two decades, the field of exoplanetary science has made breathtaking advances, both in the number of systems known and in the wealth of information. Recently we marked the 1000th extrasolar planet discovered, of which over 400 are transiting \citep{Schneider2011, Rein2012}. Such numbers are impressive in themselves but dwarfed by the 3000+ transiting exoplanet candidates \citep{tenenbaum12, fressin13, batalha13} obtained by the Kepler mission \citep{borucki96, jenkins10}, as well as the predicted tens of thousands of planets to be discovered by the GAIA mission \citep{sozzetti11}. The large number of detections suggests that planet formation is the norm in our own galaxy \citep{howard13, fressin13, batalha13, cassan12, dressing13,wright12}. Through the measurement of the planets' masses and radii we can estimate their bulk properties and get a first insight into their compositions and potential formation histories \citep{valencia13,adams08, Grasset2009,buchhave12}. To take this characterisation work to the next level, we must gain an understanding of the planet's chemical composition. The best way to probe their chemical composition is through the study of their atmospheres. For transiting planets this is feasible when the planet transits its host star in our line of sight. This allows some of the stellar light to shine through the terminator region of the planet (transmission spectroscopy). Similarly, when the star eclipses the planet (i.e. it passes behind its host star in our line of sight) we can measure the flux difference resulting from the planet's dayside emissions (emission spectroscopy).  In the last decade, a large body of work has accumulated on the atmospheric spectroscopy of transiting extrasolar planets   \citep[e.g.][]{beaulieu10,beaulieu11, charbonneau08, brogi12, bean11, swain08,swain08b, swain09a, crouzet12, deming13, grillmair08,thatte10, tinetti07, pont08, swain12, knutson11, sing11, tinetti10, mooij12,bean11b, stevenson10} also see \citet{tinetti13} for a comprehensive review. 

Given these large and ever increasing numbers of detections, it is important to understand which of these systems lend themselves to be characterised further by the use of transmission and emission spectroscopy. In the light of the mission concept The Exoplanet Characterisation Observatory, which has been studied by the European Space Agency as one of the M3 mission candidates, this question becomes critical. In this paper, we aim to quantify the number, as well as the time required to characterise spectroscopically the transiting extrasolar planets known to date. An overview of our results with examples for specific cases is given here with the full results available online ({http://www.ucl.ac.uk/exoplanets/echotargetlist/}). 

\subsection{\echo}
In the frame of ESA's Cosmic Vision programme, \echo{ }has been studied as a medium-sized M3 mission candidate for launch in the 2022 - 2024 timeframe\footnote{http://sci.esa.int/echo/} \citep{tinetti12, Tinetti2014}. During the `Phase-A study' \echo{ }has been designed as a 1 metre class telescope, passively cooled to $\sim$50 K and orbiting around the second Lagrangian Point (L2). The baseline for the payload consists of four integrated spectrographs providing continuous spectral coverage from 0.5 to 11$\mu$m (goal 0.4 to 16$\mu$m) at a resolving power ranging from R $\sim$ 300 ($\lambda < 5\mu m$) to 30 ($\lambda > 5\mu m$). For a detailed description of the telescope and payload design, we refer the reader to the literature \citep{puig12b,puig12,tinetti12, Tinetti2014, swinyard12, eccleston12, reess12, adriani12, zapata12, pascale12, focardi12, tessenyi12payload,waldmann13}.

The \echo{ }science case can be best achieved by splitting the mission lifetime into three surveys, where the instrument capabilties are optimally suited to address different classes of question. The studied targets range from super-Earth to gas giant, temperate to very hot and stellar classes M to F. The aims of these three tiers described in the \echo{ }Assessment Study Report\footnote{\label{fn:YB}http://sci.esa.int/echo/53446-echo-yellow-book/} are as follows;
\begin{itemize}
\item \textbf{Chemical Census}: Statistically complete sample to explore the key atmospheric features: albedo, bulk thermal properties, most abundant atomic and molecular species, clouds.
\item \textbf{Origin}: Addresses the question of the origin of exoplanet diversity by enabling the retrieval of the vertical thermal profiles and molecular abundances, including key and trace gases.
\item \textbf{Rosetta Stone}: Benchmark cases to get insight into the key classes of planets. This tier will provide high signal-to-noise observations yielding very refined molecular abundances, chemical gradients and atmospheric structure. Spatial and temporal resolution will enable the study of weather and climate.
\end{itemize}

The spectral resolving power (R) and signal-to-noise (SNR) target for each mission is shown in table \ref{tab:survey-res}.

\begin{table}[h]
    \begin{tabular}{l|llll}
    Survey Name     & SNR Target & R
    ($\lambda$ $<$ 5 $\mu$m) & R($\lambda$ $>$ 5 $\mu$m) & Target No. Planets\\ \hline
    Chemical Census & 5          & 50        & 30        & $>150$    \\
    Origin          & 10         & 100       & 30        & 50 -100     \\
    Rosetta Stone   & 20         & 300       & 30        & 10 - 20        \\
    \end{tabular}
    \caption{The spectral resolving power ($R=\lambda/\Delta \lambda$) and SNR requirements of each survey mode. The SNR target is the average in a chosen spectral element (see \S \ref{sec:snr-calc}). Target number of planets refers to the number of planets expected to be observed in each mode by the mission}
    \label{tab:survey-res}
\end{table}

\section{Method}
\begin{figure}[phtb]
  \caption{Simplified flow chart of the workings of ETLOS and how it links with the Open Exoplanet Catalogue, OECPy and EChOSim. OECPy loads the catalogue values, performs the necessary calculations and assumptions and then creates a planet object containing them. ETLOS takes this planet object and calls the \echo{ }end-to-end simulator (\echosim) that simulates the observation. ETLOS then interprets and plots the results}
  \centering
    \includegraphics[width=0.8\textwidth]{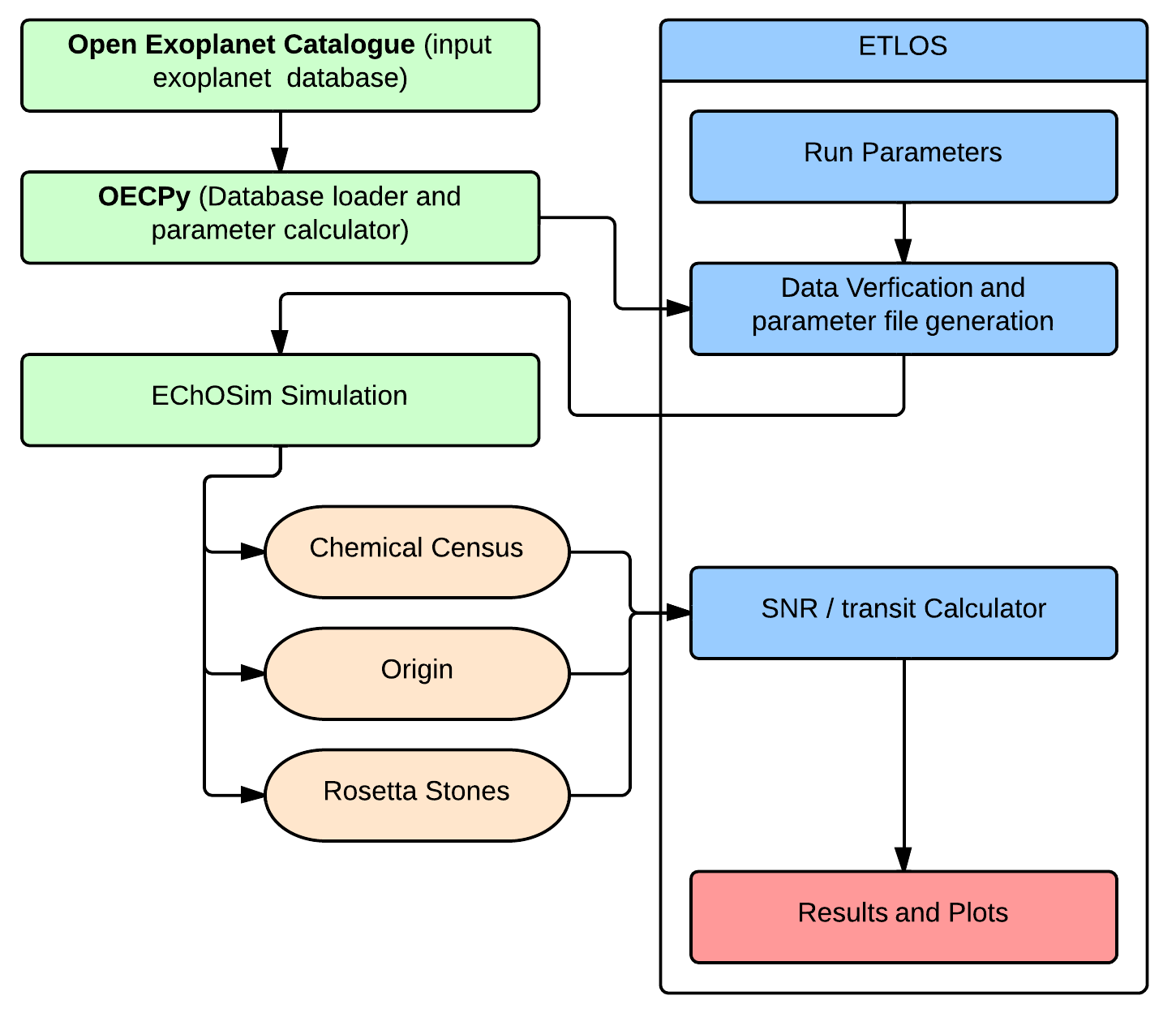}
\label{fig:etlos-flowchart}
\end{figure}

Given the instrument characteristics and the transiting planets known today, we have developed a series of models and simulators to assess the \echo{ }capabilities and optimise the science return of the mission. An overview of this generation process is shown in Fig. \ref{fig:etlos-flowchart} and detailed below.

\subsection{Planet Catalogue and OECPy}
We adopted the Open Exoplanet Catalogue (OEC, \citet{Rein2012}) for the exoplanet database which generally cites original papers as sources and is kept up-to-date as an open source community project. We have further verified most targets using SIMBAD, the 2MASS catalogue and exoplanet.eu \citep{Schneider2011} where appropriate.

To facilitate the \echo{ }Target List Simulator (ETLOS) we developed the Open Exoplanet Catalogue Python Interface (OECPy, \citet[][in prep]{Varley2014OEC}) to load the exoplanet database into ETLOS. 

OECPy is a general package for exoplanet research which retrieves exoplanet parameters from databases and performs common equations. It is the package where most of our assumptions (such as planetary classification) are handled, along with the calculations of orbital parameters and estimations for any missing values. OECPy is freely available today\footnote{https://github.com/ryanvarley/open-exoplanet-catalogue-python}.

\subsection{\echosim}
\echosim{ }\citep{waldmann13,pascale13} is the \echo{ }mission end-to-end simulator. \echosim{ } implements a detailed simulation of the major observational and instrumental effects, and associated systematics. It also allows sensitivity studies for the parameters used and thus it represents a key tool in the optimisation of the instrument design. Observation and calibration strategies, data reduction pipelines and analysis tools can all be designed effectively using the realistic outputs produced by \echosim.

The simulation output closely mimics standard STSci\footnote{http://archive.stsci.edu/hst/} FITS files, allowing for a high degree of compatibility with standard astronomical data reduction routines. In addition to \echosim, we provide an observation pipeline performing the `observed' data reduction in an optimal way \citep{waldmann14}.

\subsection{\echo{ }Target List Simulator (ETLOS)}
ETLOS works by generating the parameter files (containing the planet system information and simulation run conditions) and then simulating the observations at the spectral resolving power for each observing tier using \echosim{ }(version 3.0) and the associated observation pipeline (which analyses the \echosim{ }generated data to obtain transmission/emission spectra). \echosim{ }outputs spectra as ascii files which are used to calculate the signal to noise of the exoplanetary observation by ETLOS.
 
Along with the full simulations for each tier, a single transit and eclipse for each target was simulated and binned per spectrometer channel giving the total SNR of each channel individually. This offers a powerful way to assess the observability of a planet in each instrument module and is useful for studies of the albedo, thermal emission and orbital phase curves.

ETLOS is designed for the long term needs of the mission. New targets can be simulated as they are discovered and parameters can be changed to gauge the impact of instrument changes or optimised observation strategies. Changes can be made both globally, to targets meeting certain conditions and on a target by target basis making it a powerful informative tool.

Version 1.0 of the target list has been generated using ETLOS and The Open Exoplanet Catalogue\footnote{OEC Version (commit SHA-1) 305b90f (25th February 2014)}, future versions will be published online\footnote{http://www.ucl.ac.uk/exoplanets/echotargetlist/}.

\subsection{Selection and Run Conditions}
\label{sec:selection-conditions}
We were able to simulate 404 transiting targets from the catalogue as of 25th February 2014. The following conservative assumptions were used in our simulations to account for contingency.
\begin{itemize}
\item The Phase A instrument design, see \citet{Puig14} and \citet{Eccleston14}
\item The contamination due to the zodical light is a strong function of viewing direction. The background model (eqn. \ref{eq:zodi-calc}) is evaluated based on the Hubble model out to 2.5 micron, and the DIRBE model at wavelengths beyond \citep{Kelsall1998}. The value we use is 3 times this expression for all targets. This is above the average value of 2.5 (the actual value varies between 0.9 at the ecliptic poles to 8 in a small number of extreme cases).
\begin{equation}
Zodi(\lambda)= B_\lambda (5500 K)\times \text{3.5E-14} + B_\lambda (270 K)\times \text{3.58E-8} \times I
\label{eq:zodi-calc}
\end{equation}
in units of $W/m^2/sr/m$, where $B_\lambda(T)$ is Planck's law written in terms of wavelength at a temperature of T $K$. 
\item No flux correlation along wavelengths is assumed. A theoretical SNR gain of up to $\sqrt{2}$ is achievable given that lightcurves are fully correlated over wavelength. 
\item Telescope jitter of  20 mas-rms at 2.8$\times$10$^{-2}$- 1 Hz and 50 mas-rms at 1 - 300 Hz  \citep[see ][]{waldmann13}
\item Circular orbits are assumed for all targets in this initial study.
\item Exotic targets with eccentricity $>0.5$, like HD 80606 b along with targets around binary stars, cannot be automatically simulated in the current iteration and need to be modelled separately. Future versions will include these targets automatically through improved calculations.
\end{itemize}

\subsection{Missing information in catalogues}
\label{sec:Missing-info}
When measured parameters such as mass and inclination are unknown, we infer them using OECPy. Our assumptions for these cases are described below.

\begin{itemize}
\item Inclination = $90\deg$
\item Planetary effective temperature ($T_{\text{pl}}$) is estimated using the equation:
\begin{equation}
T_{pl} = T_\star \left( \sqrt{\frac{(1-A)}{\epsilon}\frac{R_\star}{2a}}\right)^{1/2}
\label{eq:planet-temp}
\end{equation}
where the planetary albedo is given in table \ref{tab:temperature-assum} and a greenhouse effect contribution of $\epsilon = 0.7$ is assumed \citep{Tessenyi2012, Seager2003}.
\item Planetary mass ($M_\text{p}$) was estimated using the density of planet classes included in Table \ref{tab:type-assum} and the measured planetary radius ($R_\text{p}$).
\item The distance to the star is estimated first by using an absolute magnitude lookup table based on spectral type\footnote{http://xoomer.virgilio.it/hrtrace/Sk.htm\_SK3 from Schmid-Kaler (1982)} and then using the distance magnitude relationship (eqn. \ref{eq:distance-magnitude}) to calculate the distance. We do not correct for absorption as we 
adopt a conservative estimate.
\begin{equation}
m-M = 5 \log_{10}{d} -5
\label{eq:distance-magnitude}
\end{equation}
\item Apparent magnitude (in K band) is calculated by converting the K band magnitude using table A5 of \citet{Kenyon1995}. This is only used for distance estimation when the V band magnitude is unavailable.
\item The mean molecular weight has been estimated according to the planetary classes given in table \ref{tab:type-assum}. In particular we assumed molecular hydrogen for gaseous Jupiters and Neptunes and water vapour for super-Earths.
\end{itemize}

\begin{table}
    \begin{tabular}{l|lll}
    ~                            & Jupiter & Neptune & Super-Earth \\ \hline
    Mass ($M_\oplus$)            & $>$ 50  & $>$ 10  & $<$ 10      \\
    Radius ($R_\oplus$)        & $>$ 6   & $>$ 3   & $<$ 3       \\
    Density (g/$cm^3$)           & 1.326   & 1.638   & 4           \\
    Mean Molecular Weight (a.m.u) & 2       & 2       & 18          \\
    \end{tabular}
    \caption{Our assumptions for target values based on type. Note that we use mass to classify a planet first, if mass is missing we then use the radius. In this case the mass is estimated from the densities from \citet{Grasset2009} which are given in this table and is used in the calculation of the scale height within \echosim}
    \label{tab:type-assum}
\end{table}

\begin{table}
    \begin{tabular}{l|ll}
    Type      & T (K)   & Bond Albedo \\
    \hline
    Hot       & $>$ 700 & 0.1         \\
    Warm      & $>$ 350 & 0.3         \\
    Temperate & $<$ 350 & 0.3         \\
    \end{tabular}
    \caption{Planet assumptions based on planetary effective temperature}
    \label{tab:temperature-assum}
\end{table}

\subsection{SNR Calculation}
\label{sec:snr-calc}
To assess the observability of the atmospheres, we first simulated featureless transmission spectra and black body emission spectra. We then assessed the detectability of specific molecular features at different abundances for a selection of planets in transit and eclipse (see \S \ref{sec:mol-verification}). Each of the spectrographs in \echo{ }are simulated (VNIR $0.55$-$2.5 \mu m$; SWIR $2.5$-$5.0 \mu m$; MW1IR $5.0$-$8.5 \mu m$; MW2IR $8.5$-$11.0 \mu m$; LWIR $11.0$-$16.0 \mu m$). The observability of a target is assessed by taking the mean SNR of the SWIR and two MWIR channels (covering 2.5-11$\mu m$). The next section shows \echosim{ }simulated spectra with varying molecular species simulated using the number of transits determined by the original (featureless) calculations.

\subsection{Molecular observability and SNR Validation}
\label{sec:mol-verification}
To verify our SNR choices for the different tiers are indeed optimal, we ran three planets (GJ 1214b, GJ 436b, HD 189733b) in transmission for $CO$, $CO_2$, $H_2O$, $CH_4$ at abundances of $10^{-3}$, $10^{-5}$, $10^{-7}$ and three planets (55 Cnc e, GJ 436b, HD 189733b) in emission with the same compositions and abundances\footnote{Only $CO_2$ and $H_2O$ were ran for 55 Cnc b as the other cases are unrealistic given its temperature}. The spectral files were generated using TAU \citep{Hollis2013} for transmission and as described by \citet{Tessenyi2013} for emission.

The simulations were ran using the number of transits calculated in our original featureless simulations.
  
\section{Results}
We find that as of 25th February 2014, 173 planets are observable in the \echo{ }Chemical Census observation tier within the proposed mission lifetime of 4 years in transit or eclipse of which 162 are observable in both.  In Origin 165 targets are observable in transit or eclipse with 148 in both. In Rossetta Stone 132 targets are observable in transit or eclipse with 78 in both. We note the recent discovery of over 700 planets \citep{2014Lissauer, 2014Rowe} are not included in our results and will be added in a future version.

\begin{figure}[p]
  \caption{Example of the signal-to-noise plots generated by ETLOS (through \echosim ) for Gilese 3470 b in transmission (left) and Gilese 436 b emission (right). The upper plots show the SNR per bin at the number of transits required for each \echo{ }tier. The lower plots demonstrate the differences of each tier by showing a simulated molecular case for each planet (offset for clarity) at the number of transits per mode given in the upper plots. Gliese 3470 b is simulated using a composition of 8E-3 $H_2O$, 4E-3 $CH_4$, 2E-3 $CO$, 2E-5 $CO_2$, 1E-4 $NH_3$ generated by TauRex \citep{taurex2014}. Depending on the radius and temperature ratios of the planet and the star, some spectral bands are more appropriate where others may be less informative. By covering a broad wavelength range we can cover a large range of planets from Jupiter to super-Earth, hot to temperate.}
  \centering
  \begin{tabular}{cc}
    \includegraphics[width=0.5\textwidth]{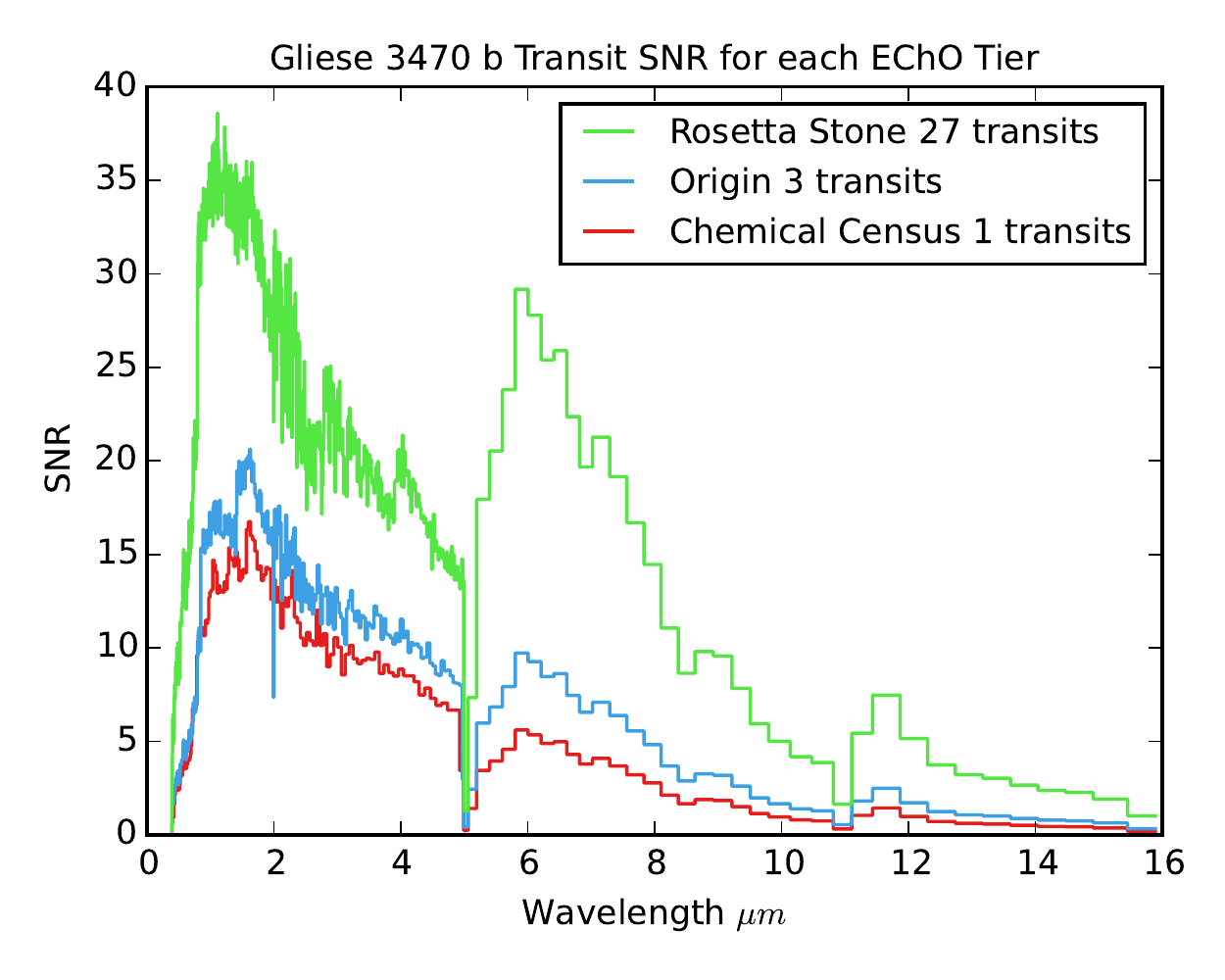}&
    \includegraphics[width=0.5\textwidth]{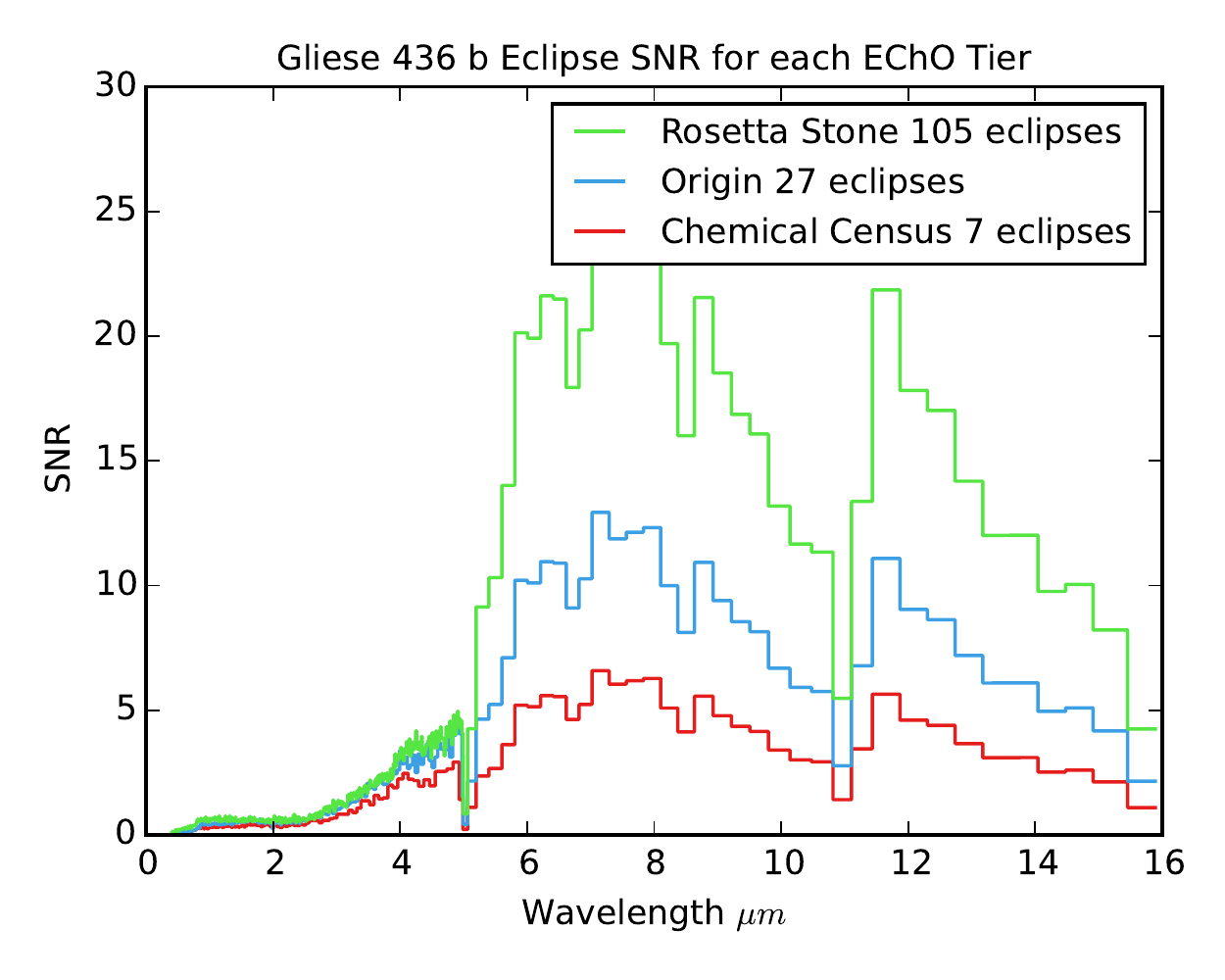}\\
    \includegraphics[width=0.5\textwidth]{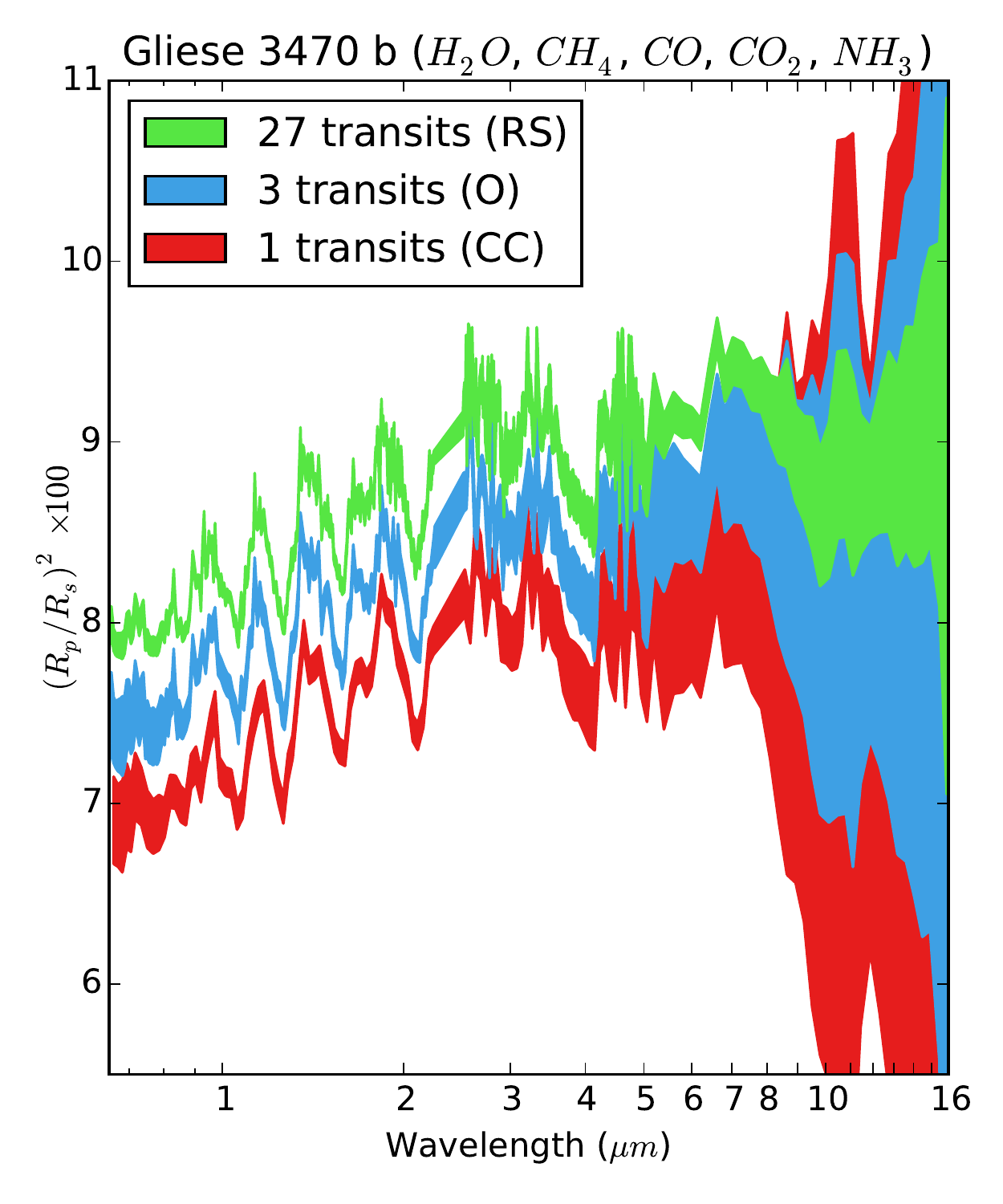}&
    \includegraphics[width=0.5\textwidth]{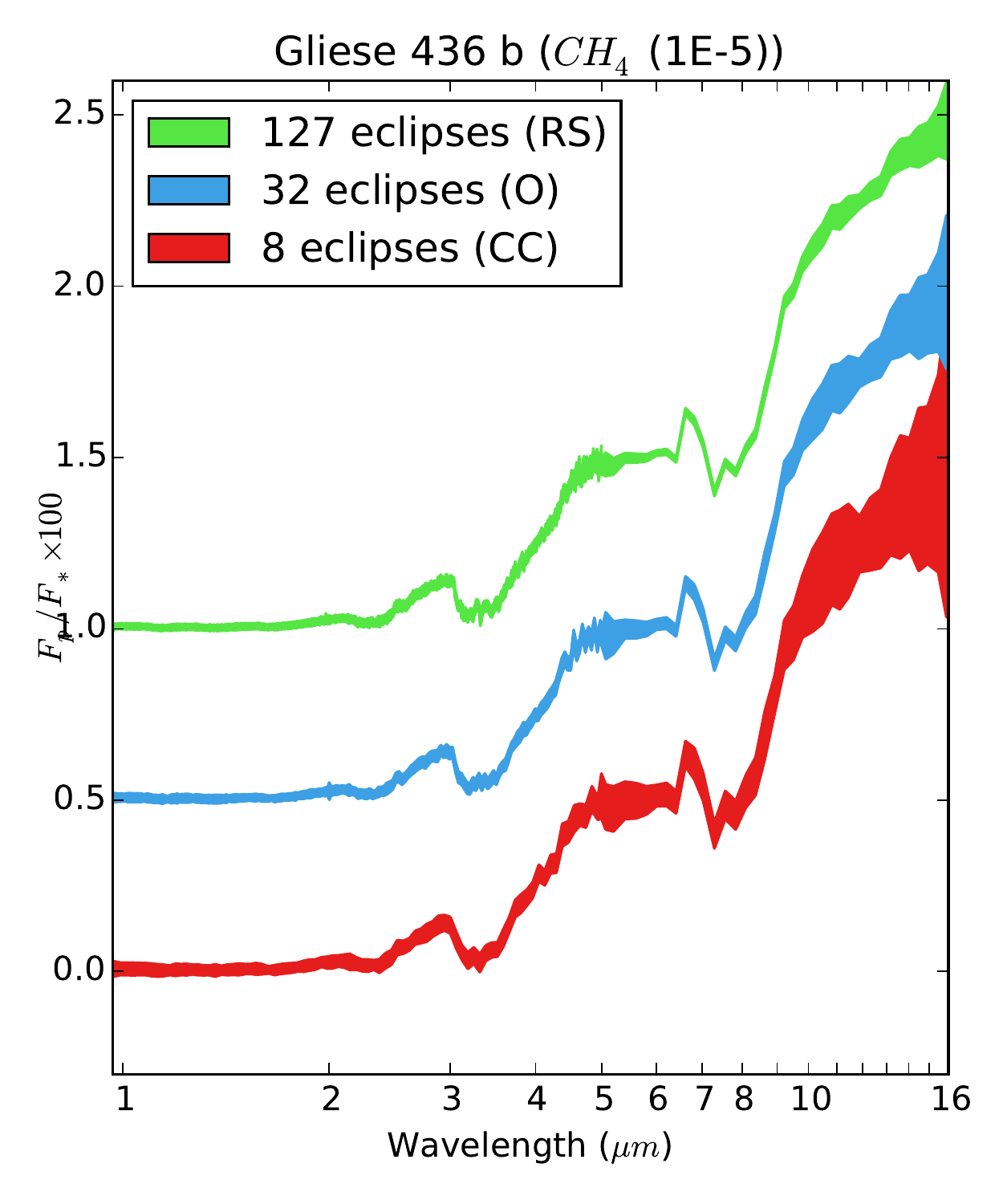}\\
    \end{tabular}
\label{fig:example-snr}
\end{figure}

\begin{figure}[p]
  \caption{Example of the per instrument module signal-to-noise plots generated by ETLOS (through \echosim ) for Gilese 3470 b in transmission (left) and Gilese 436 b emission (right). The plots show the SNR of each instrument module as a single bin for a single transit and eclipse.}
  \centering
  \begin{tabular}{cc}
    \includegraphics[width=0.5\textwidth]{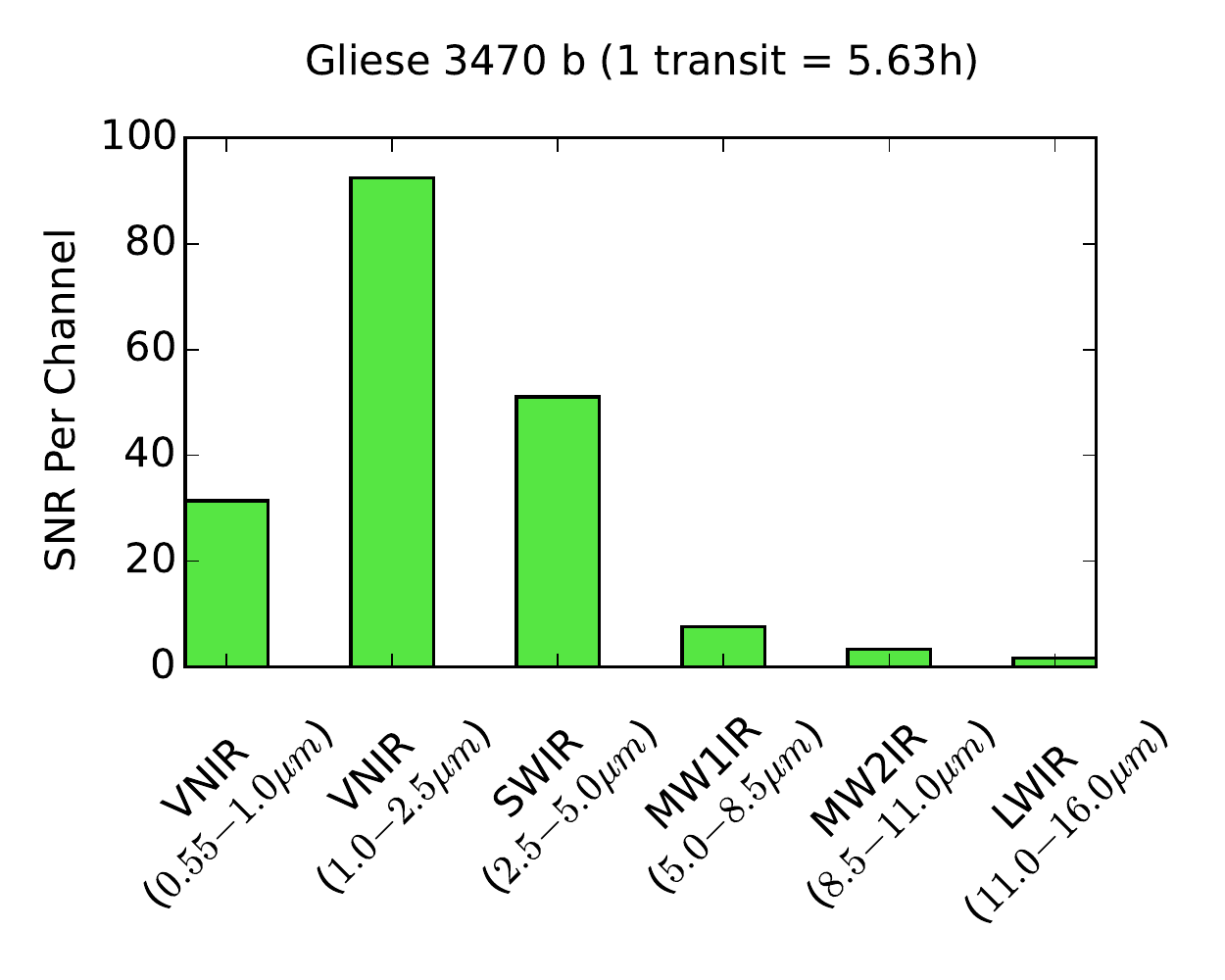}&
    \includegraphics[width=0.5\textwidth]{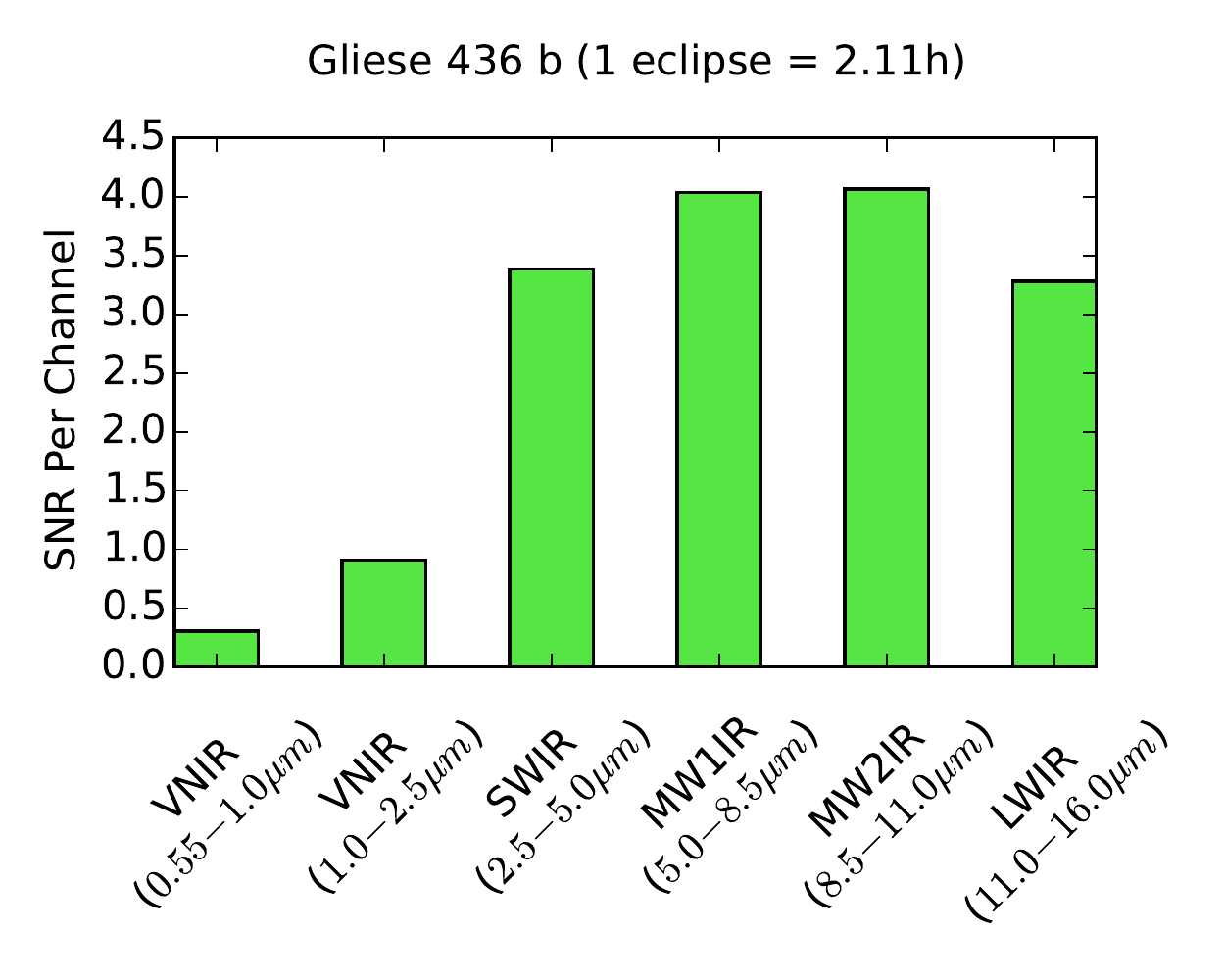}\\
    \end{tabular}
\label{fig:example-snr-bulk}
\end{figure}

\begin{figure}[p]
  \caption{Examples of the molecular cases simulated. The plots show the types of features detectable in each tier with Rosseta Stone being able to constrain features in models with much lower abundances. See {http://www.ucl.ac.uk/exoplanets/echotargetlist/} for the other simulations.}
  \centering
    \begin{tabular}{cc}
    \includegraphics[width=0.5\textwidth]{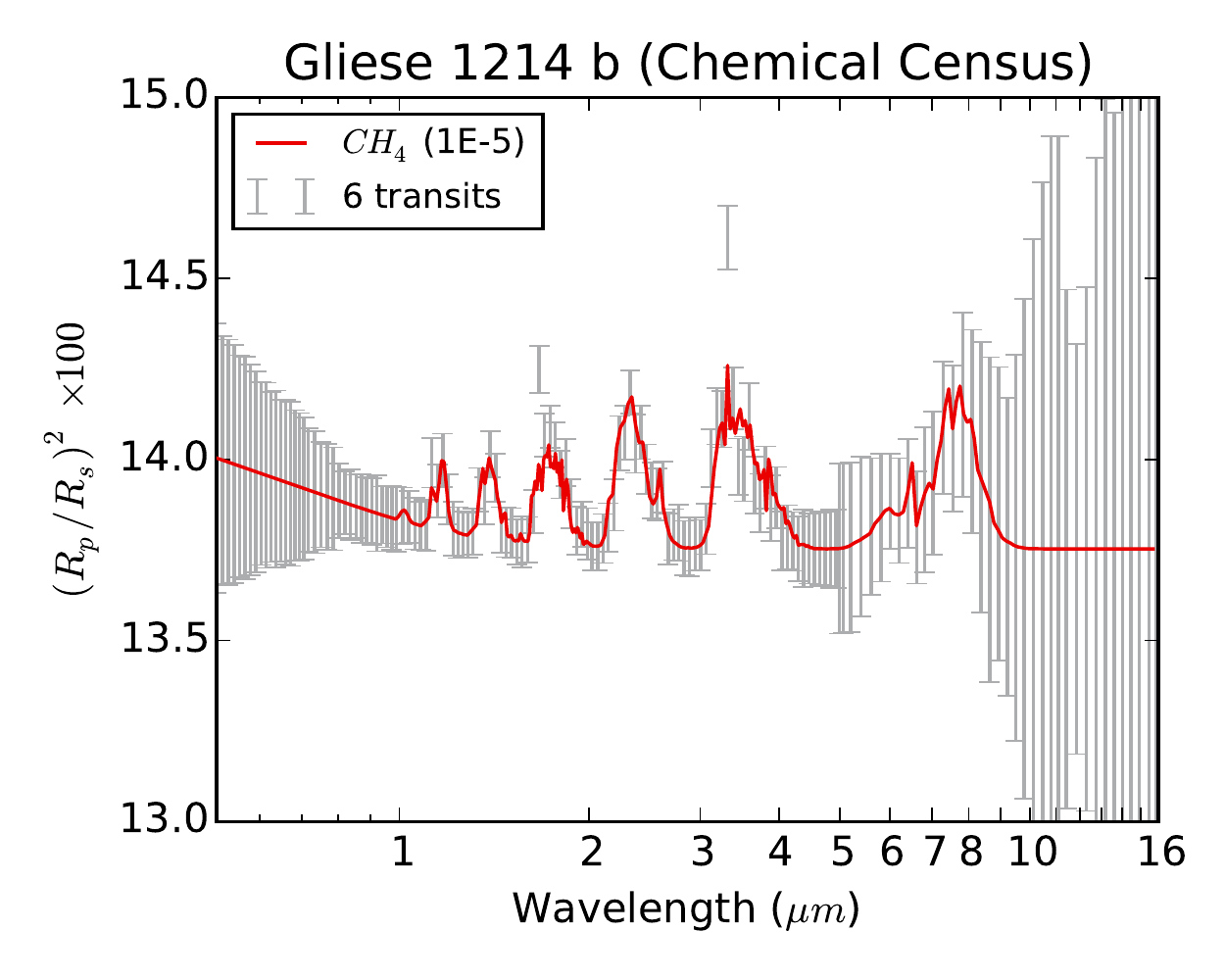}&
    \includegraphics[width=0.5\textwidth]{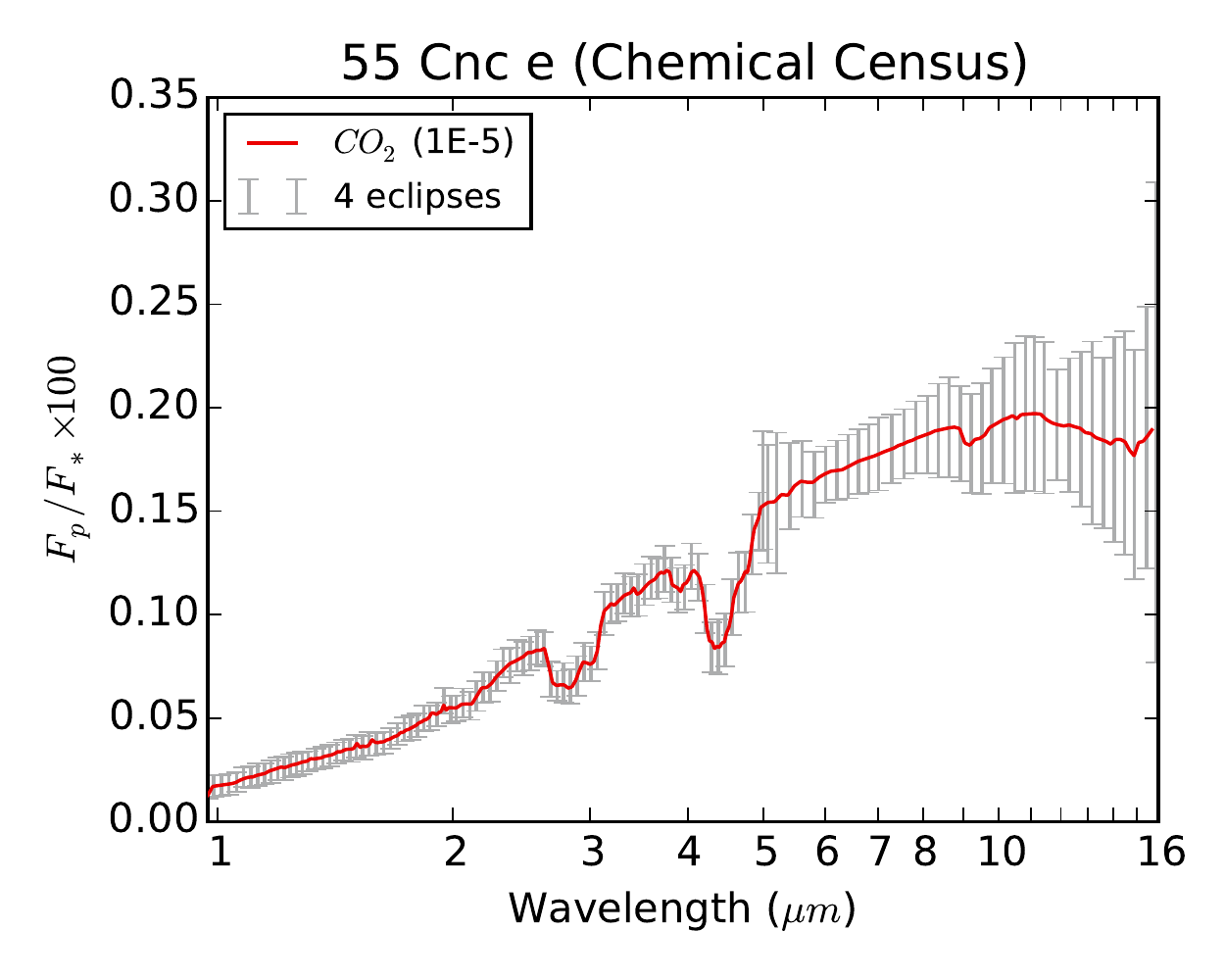}\\
    \includegraphics[width=0.5\textwidth]{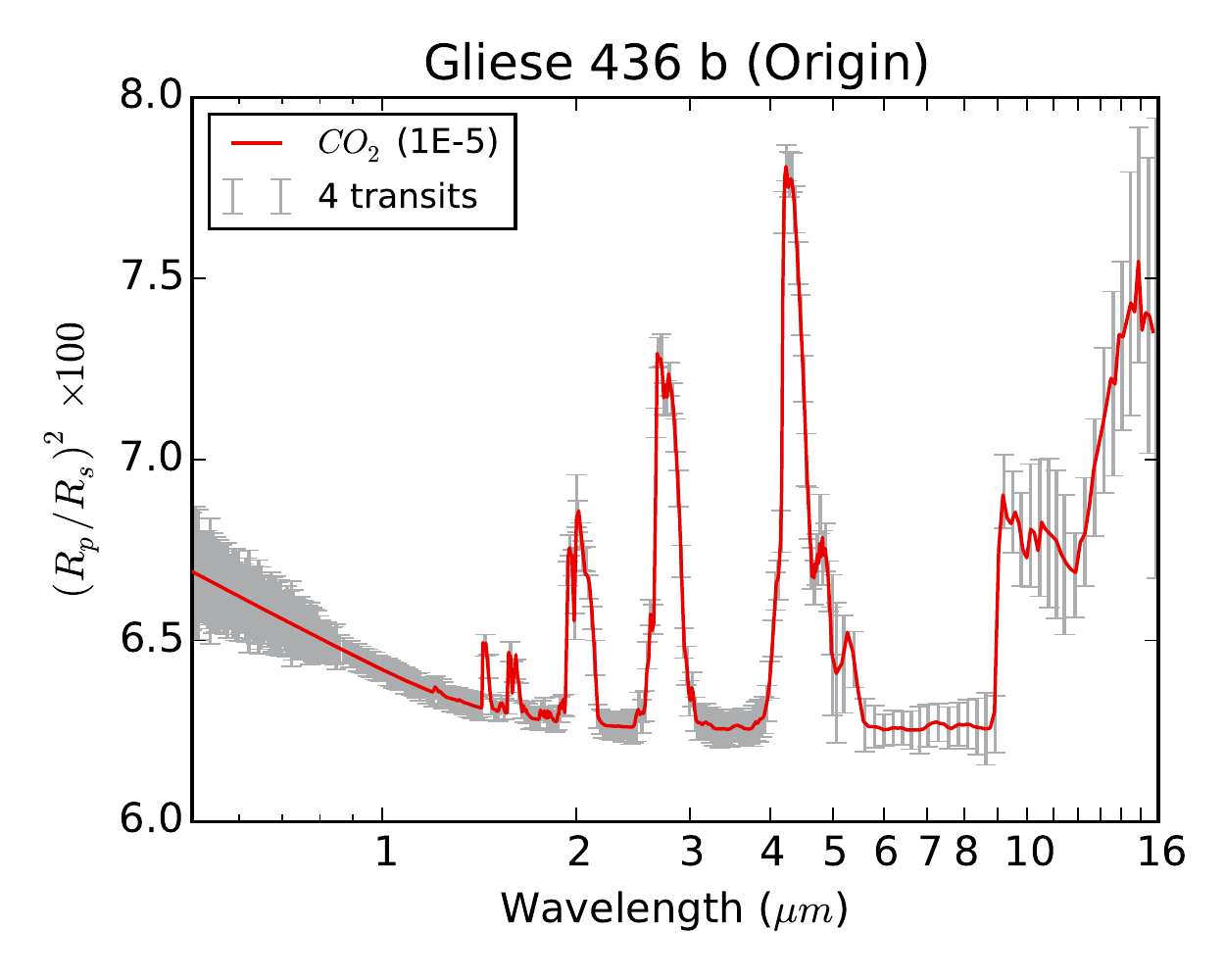}&
    \includegraphics[width=0.5\textwidth]{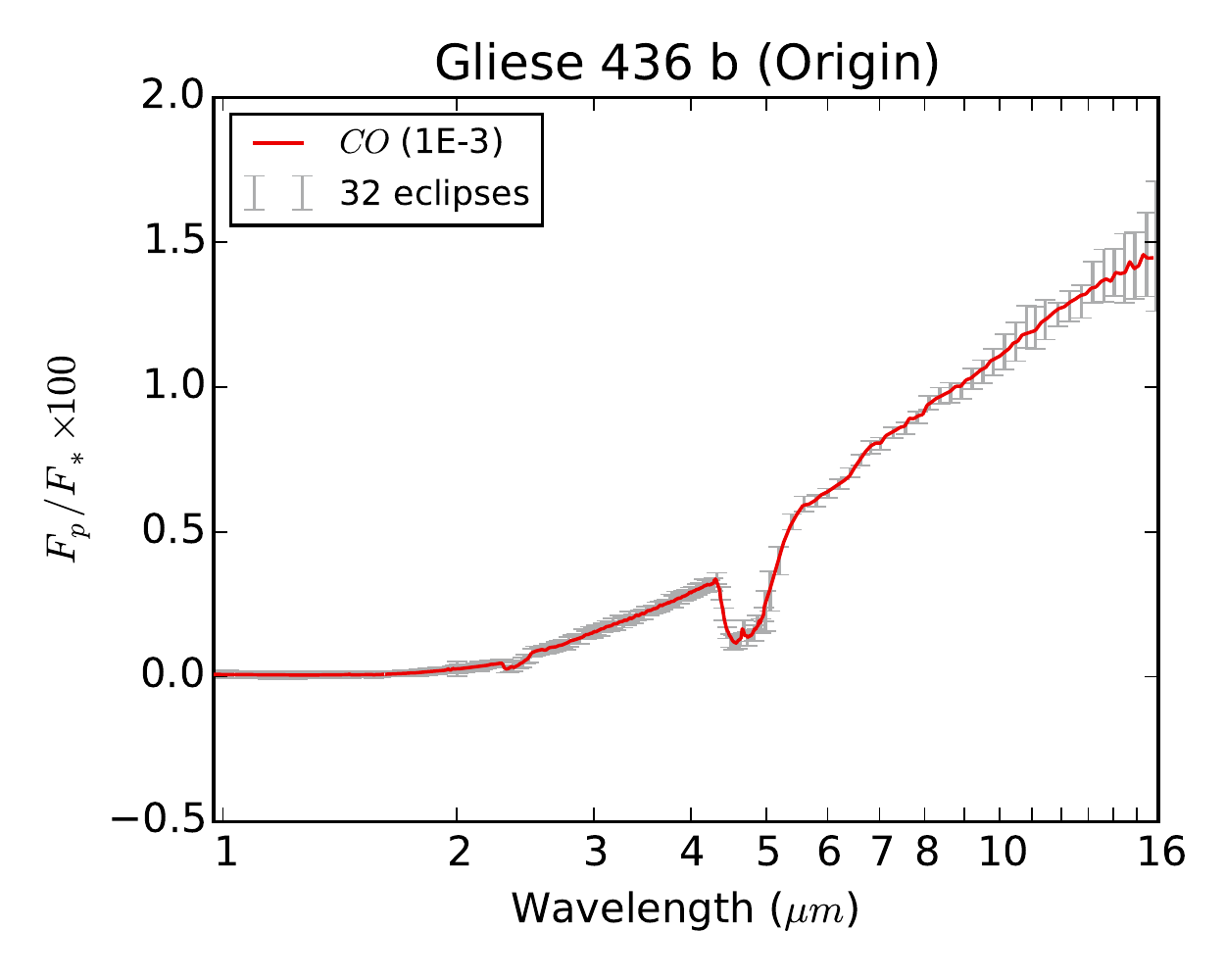}\\
    \includegraphics[width=0.5\textwidth]{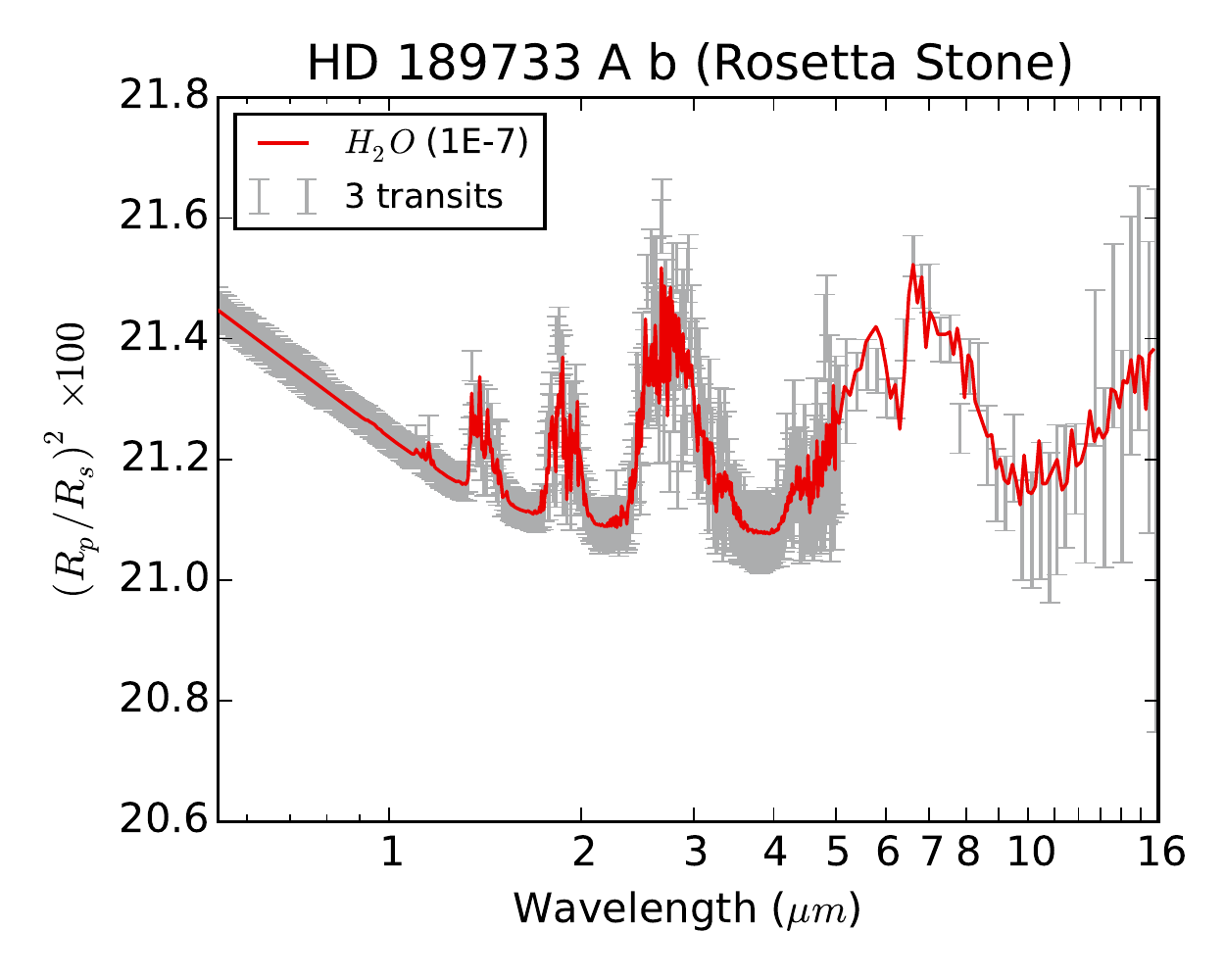}&
    \includegraphics[width=0.5\textwidth]{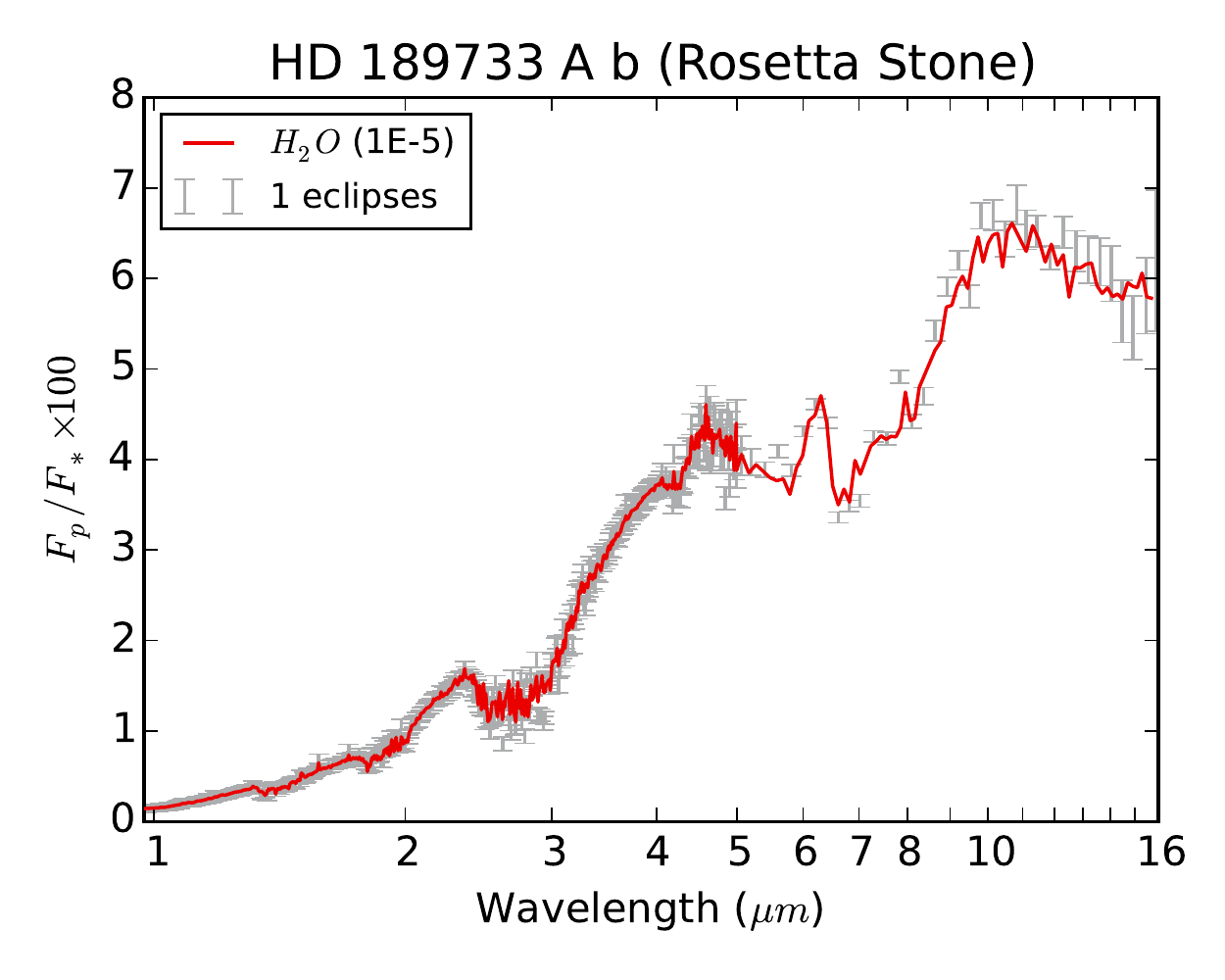}\\
    \end{tabular}
\label{fig:mol-spectra-plots}
\end{figure}

We generated three types of plot per target to show the signal-to-noise of each target with wavelength: 
\begin{itemize}
\item SNR of the cumulative observations of transits required to fulfil the requirements for each observing tier (from table \ref{tab:survey-res}), Fig. \ref{fig:example-snr}
\item Multiple broadband photometry for a single transit and eclipse, Fig. \ref{fig:example-snr-bulk}
\item For a subset of targets simulations of transit and eclipse spectra indicating the strength of  molecular features when the appropriate SNR and R are reached. Fig. \ref{fig:mol-spectra-plots}
\end{itemize}
In addition to the examples shown here, the plots for all observable targets are available online ({http://www.ucl.ac.uk/exoplanets/echotargetlist/}).

Our simulations of molecular features show that even for low abundances (e.g. mixing ratios  $<10^{-5}$) the Chemical Census tier is sufficient to detect most of the trace gases (see also \citet{Tessenyi2013} \citet[][in prep.]{Tessenyi2014}) with Origin and Rossetta Stone tiers being increasingly able to constrain spectral features.

\begin{figure}[p]
  \caption{Planetary parameter space probed today by the Chemical Census tier.}
  \centering
    \begin{tabular}{cc}
    \includegraphics[width=0.5\textwidth]{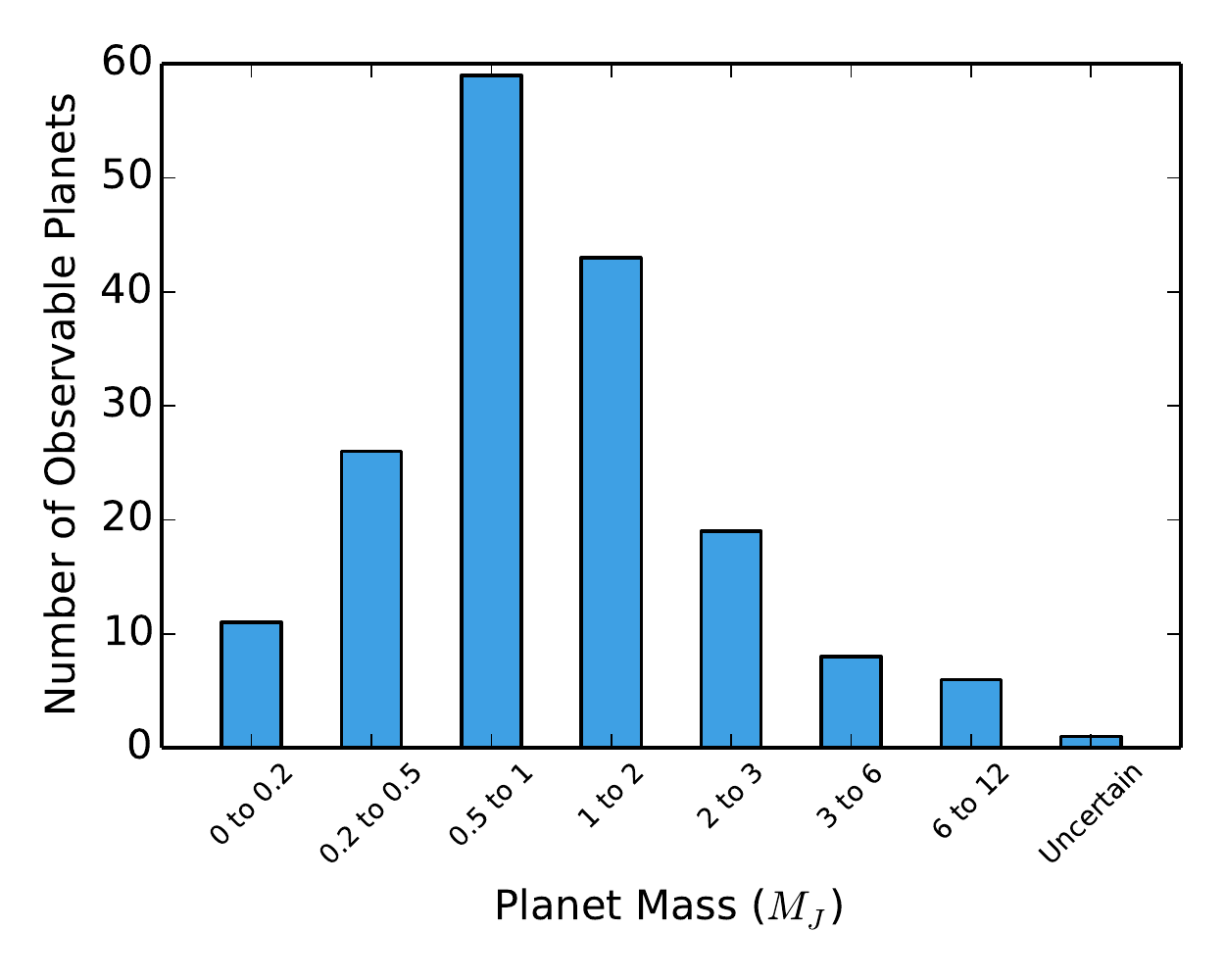}&
    \includegraphics[width=0.5\textwidth]{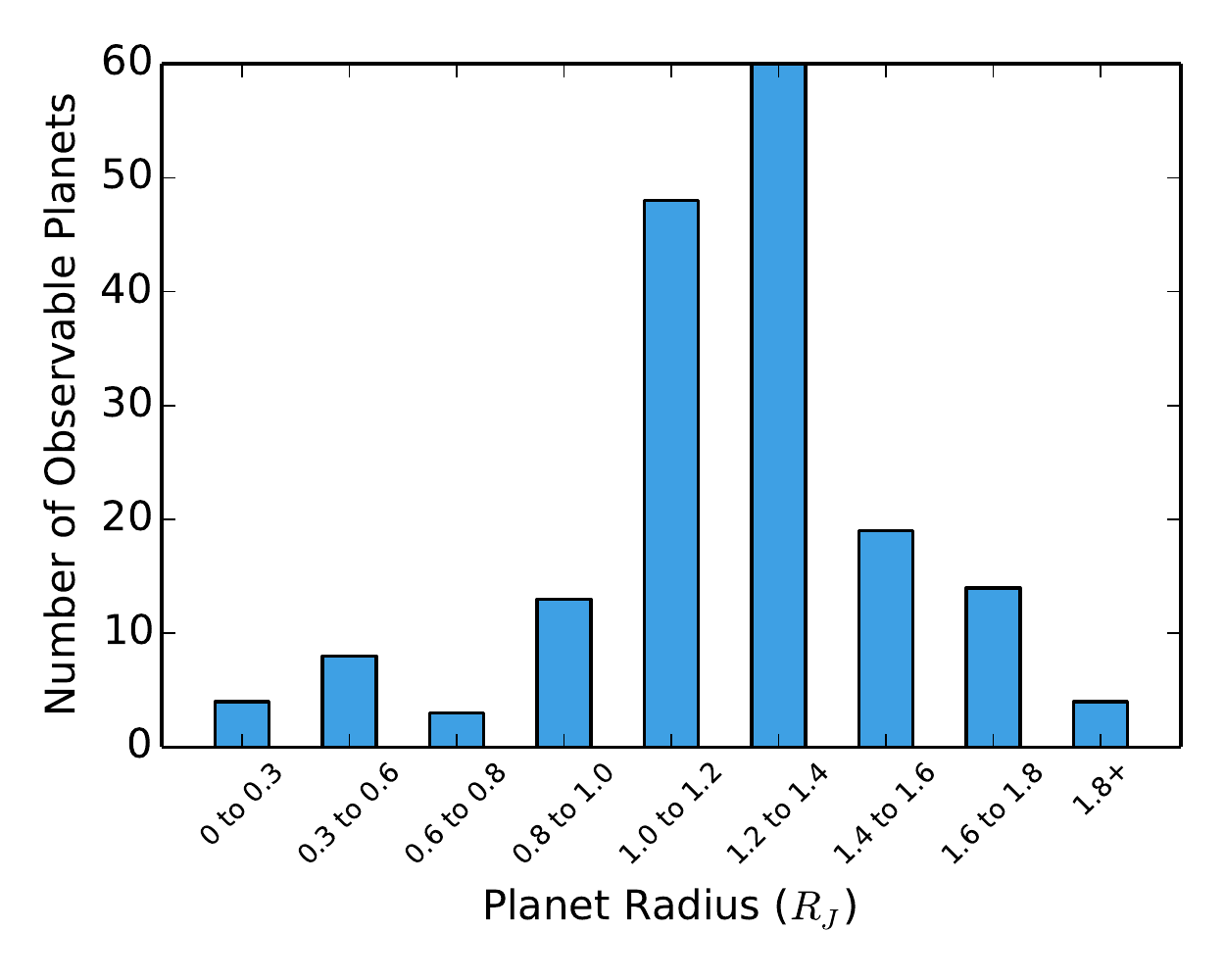}\\
    \includegraphics[width=0.5\textwidth]{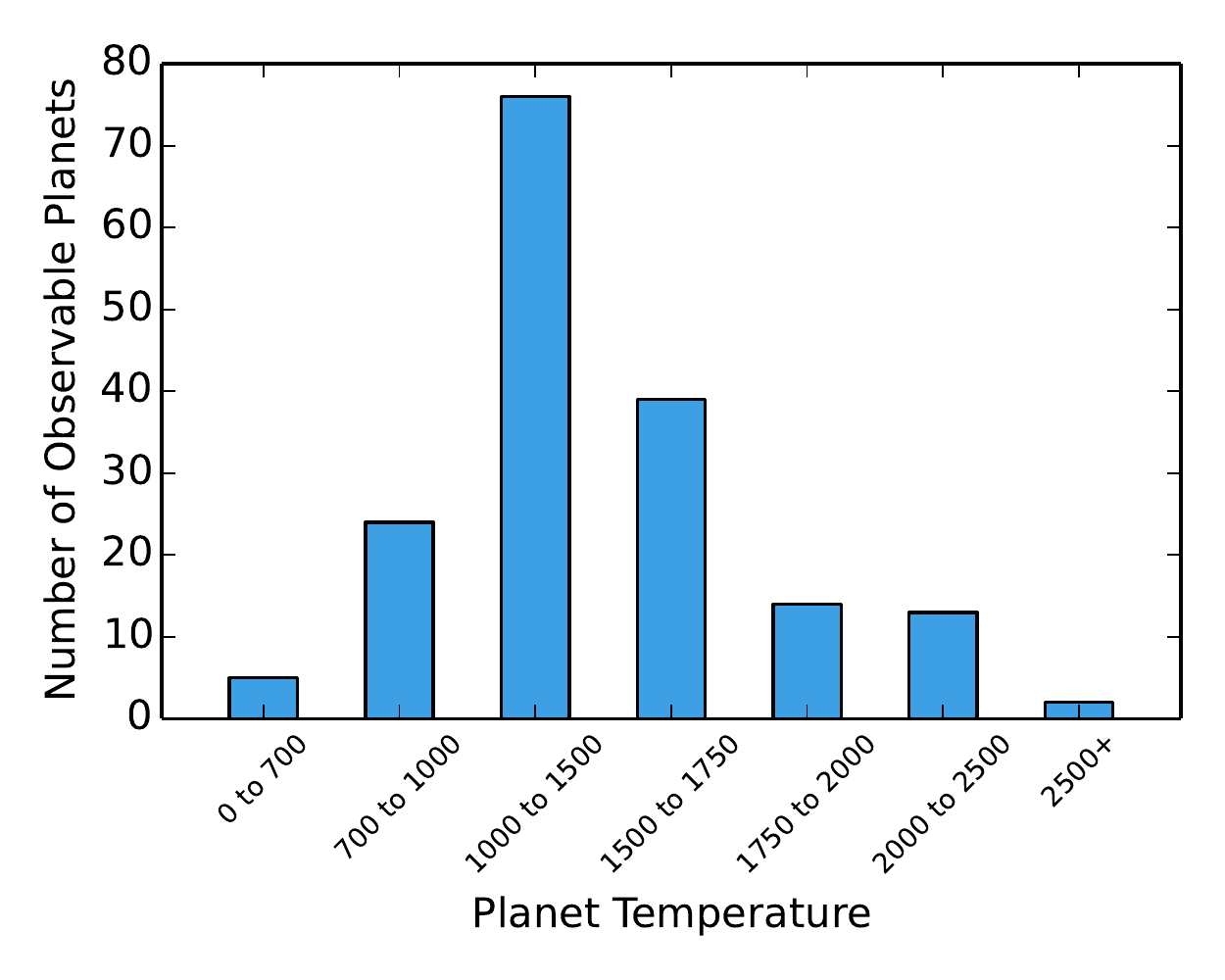}&
    \includegraphics[width=0.5\textwidth]{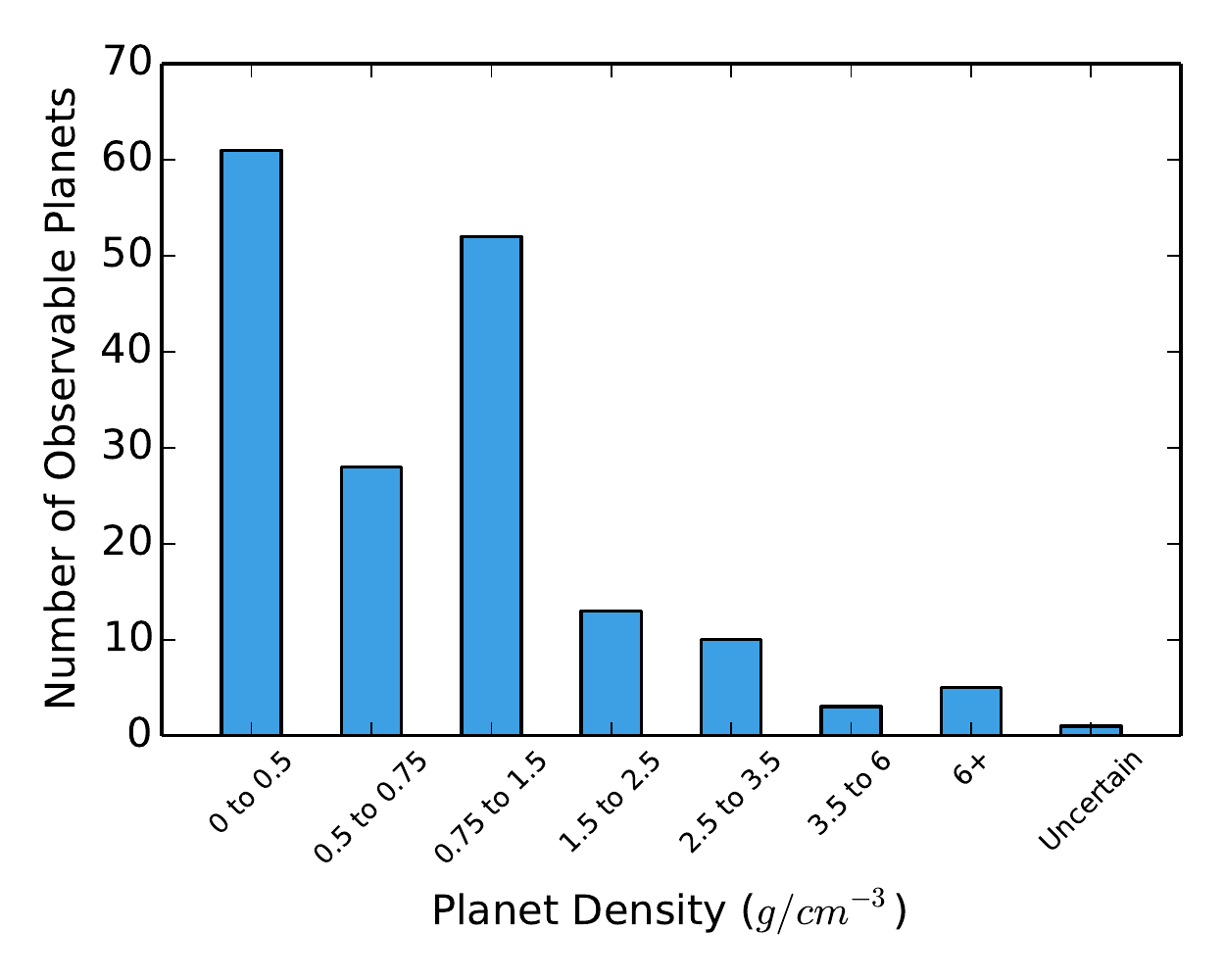}\\
    \includegraphics[width=0.5\textwidth]{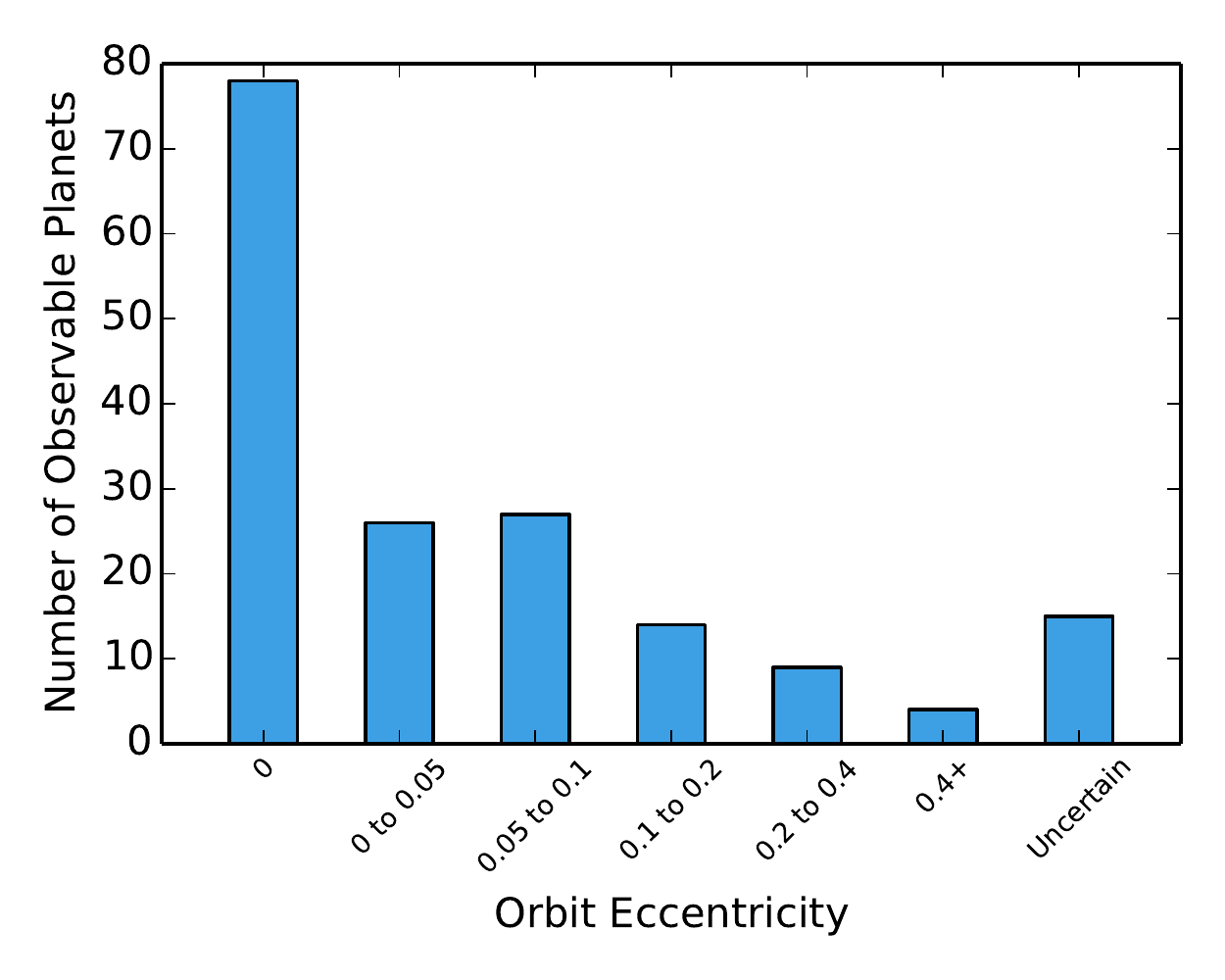}
    \end{tabular}
\label{fig:par-space-planet}
\end{figure}

\begin{figure}[p]
  \caption{Stellar parameter space probed today by the Chemical Census tier.}
  \centering
    \begin{tabular}{cc}
    \includegraphics[width=0.5\textwidth]{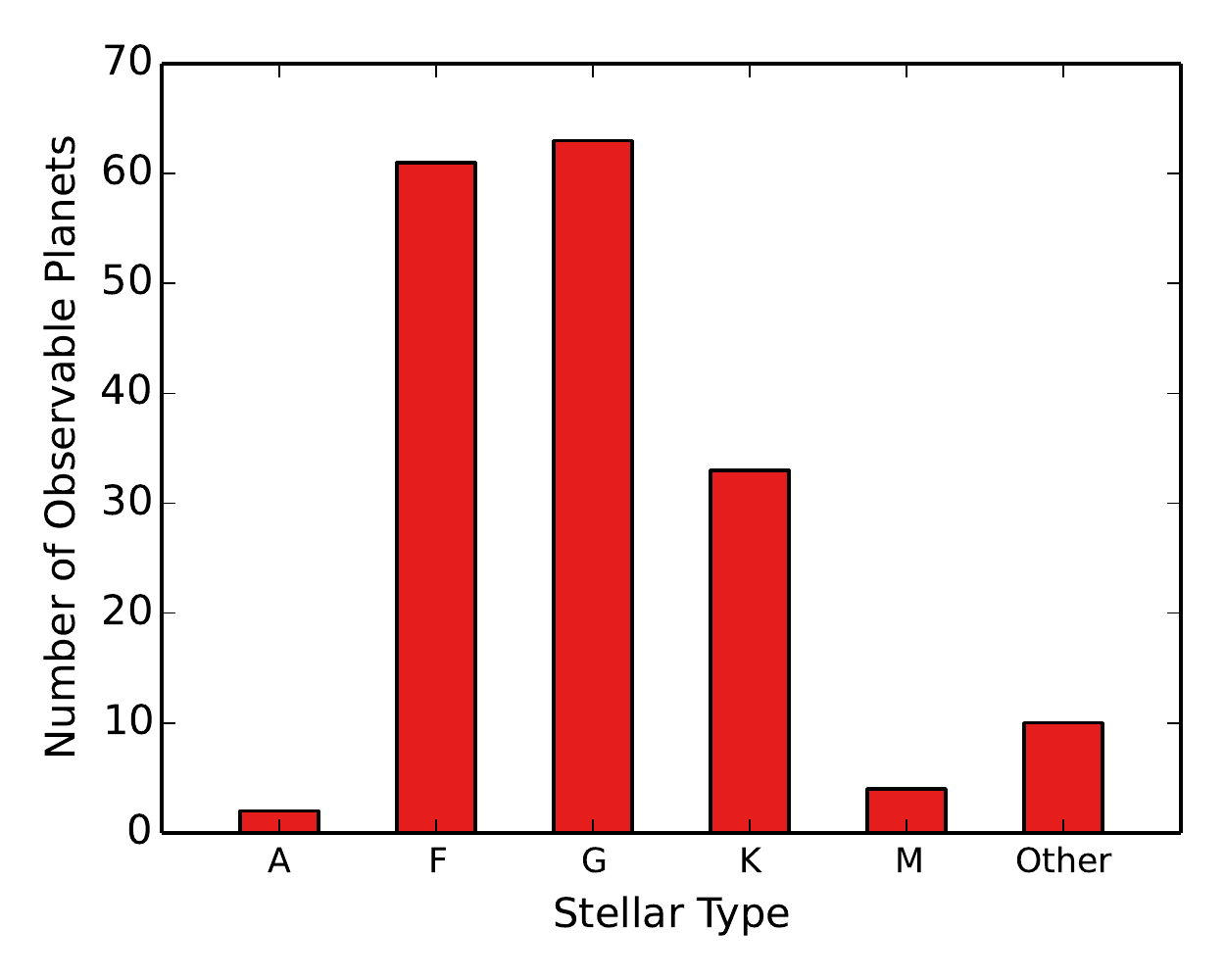}&
    \includegraphics[width=0.5\textwidth]{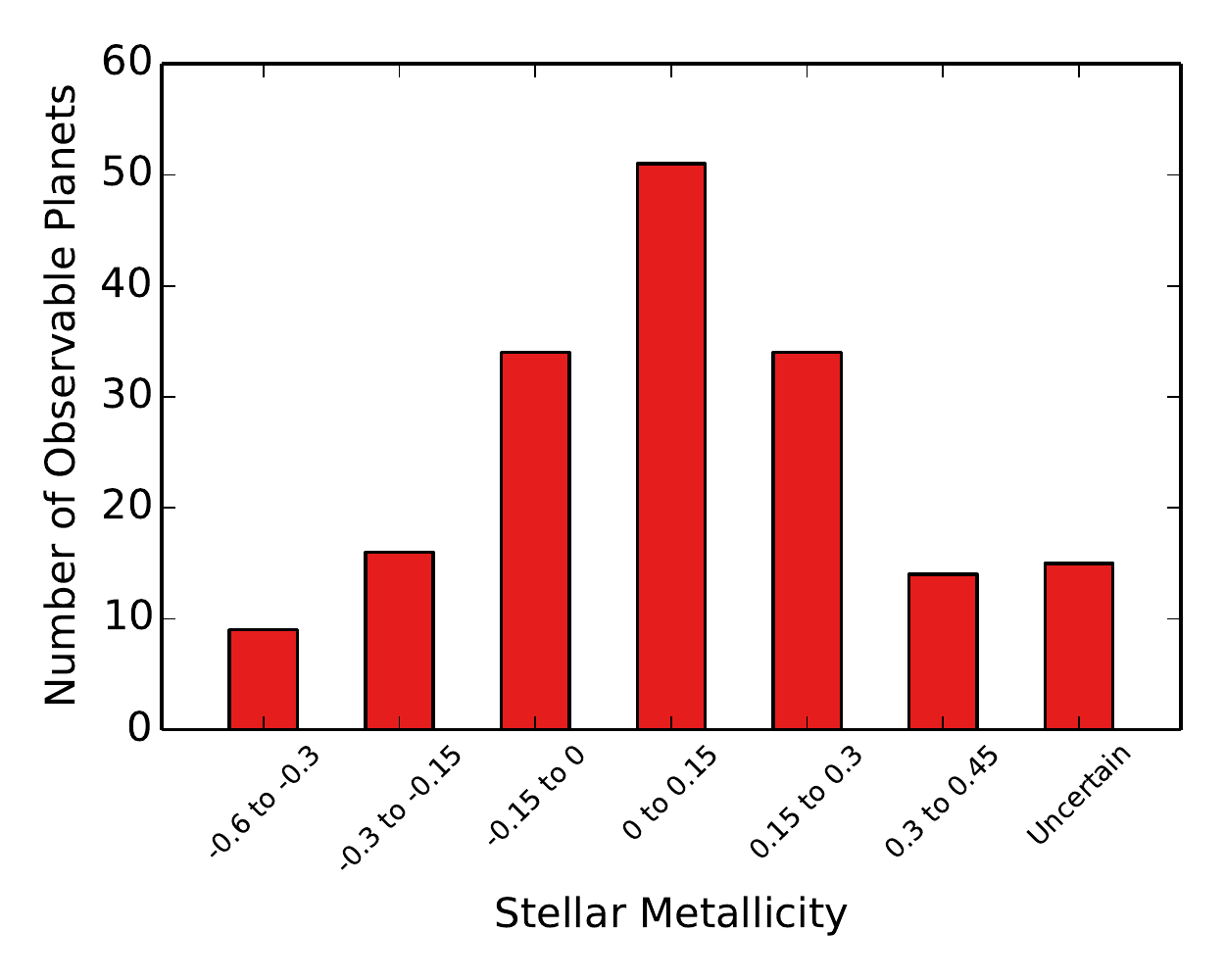}\\
    \includegraphics[width=0.5\textwidth]{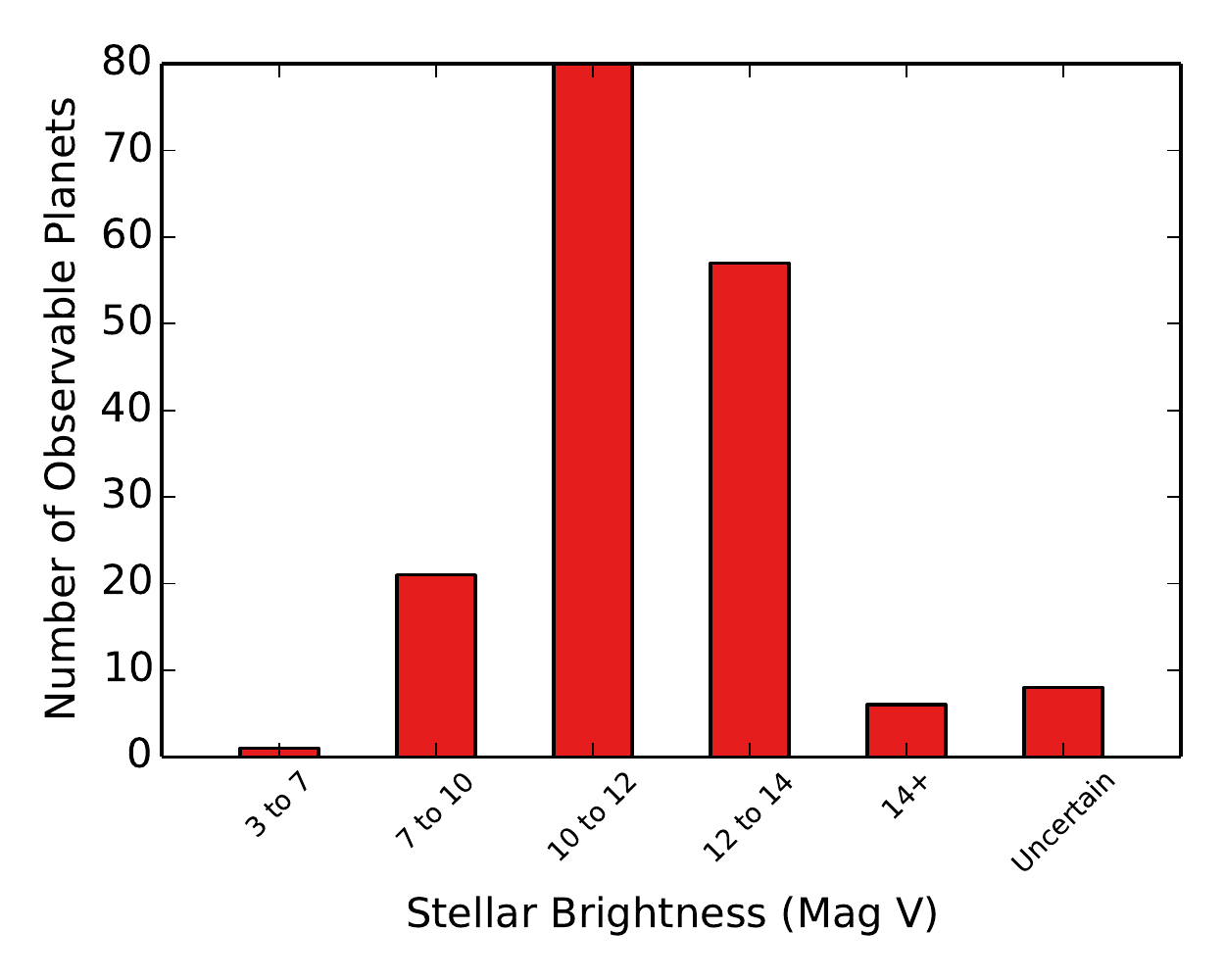}&
    \includegraphics[width=0.5\textwidth]{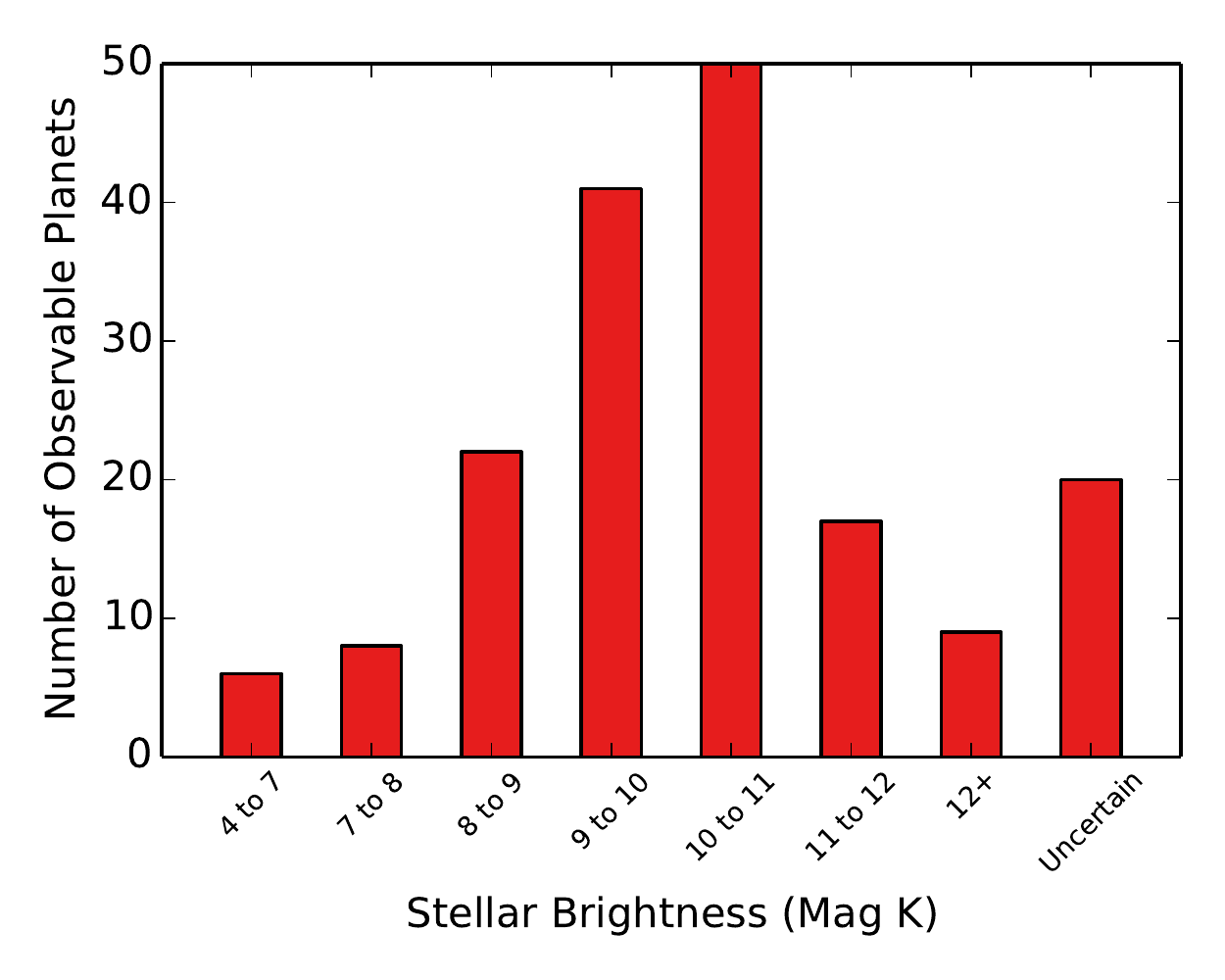}\\
    \end{tabular}
\label{fig:par-space-star}
\end{figure}

Figures \ref{fig:par-space-planet} \& \ref{fig:par-space-star} show the parameter space that \echo{ }could explore if launched today. Observable systems cover a very broad parameter space in terms of stellar types, eccentricity, temperatures and densities.

\begin{figure}[p]
  \caption{Plots showing the stellar magnitude with planetary radii for current observable population for Left: The current observable population with \echo. Right: the predicted yield of planets from the Transiting Exoplanet Survey Satellite (TESS, figure published with permission from the TESS Concept Study Report submitted in September 2012).}
  \centering
    \begin{tabular}{cc}
    \includegraphics[width=0.5\textwidth]{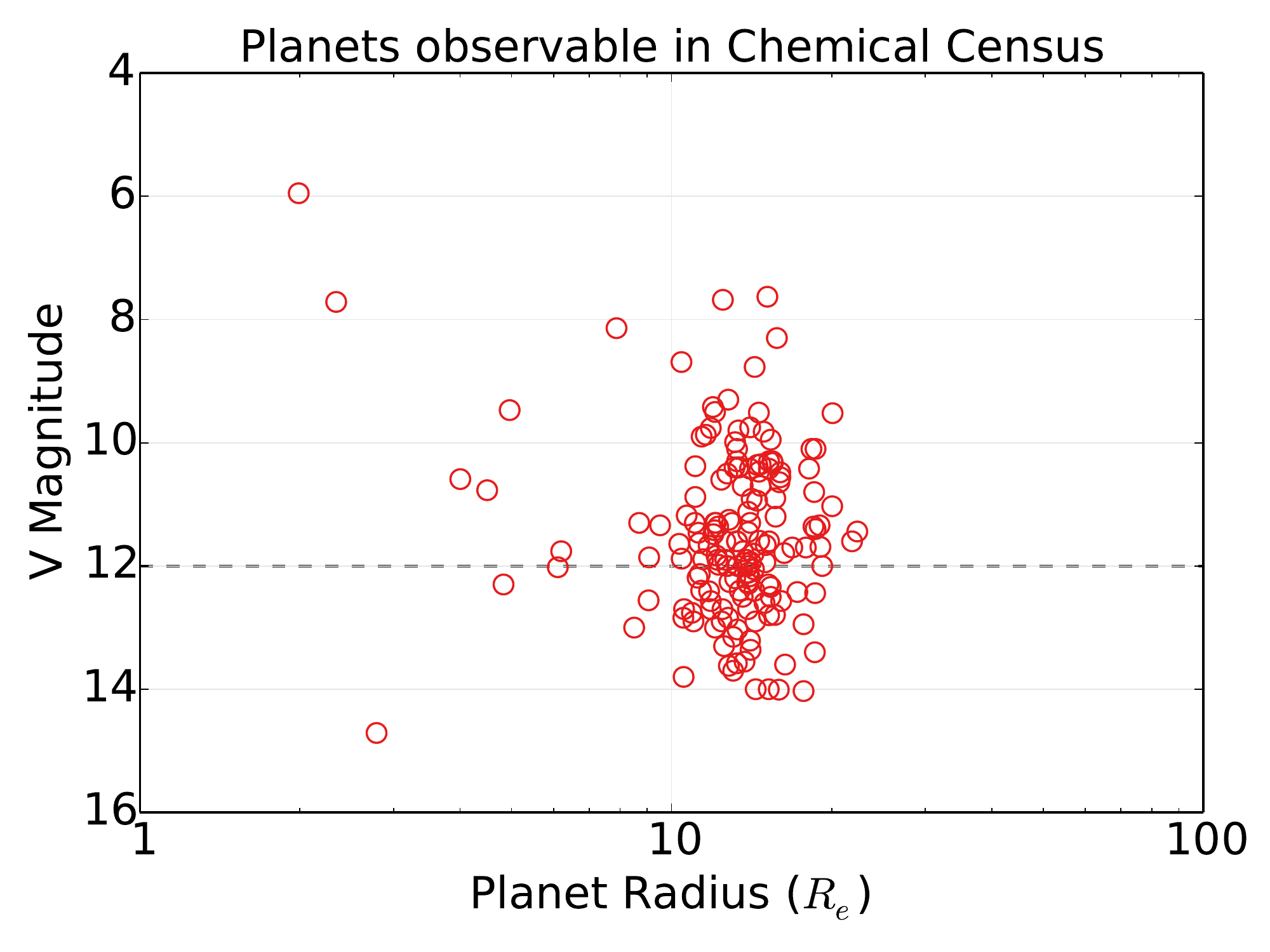}&
    \includegraphics[width=0.45\textwidth]{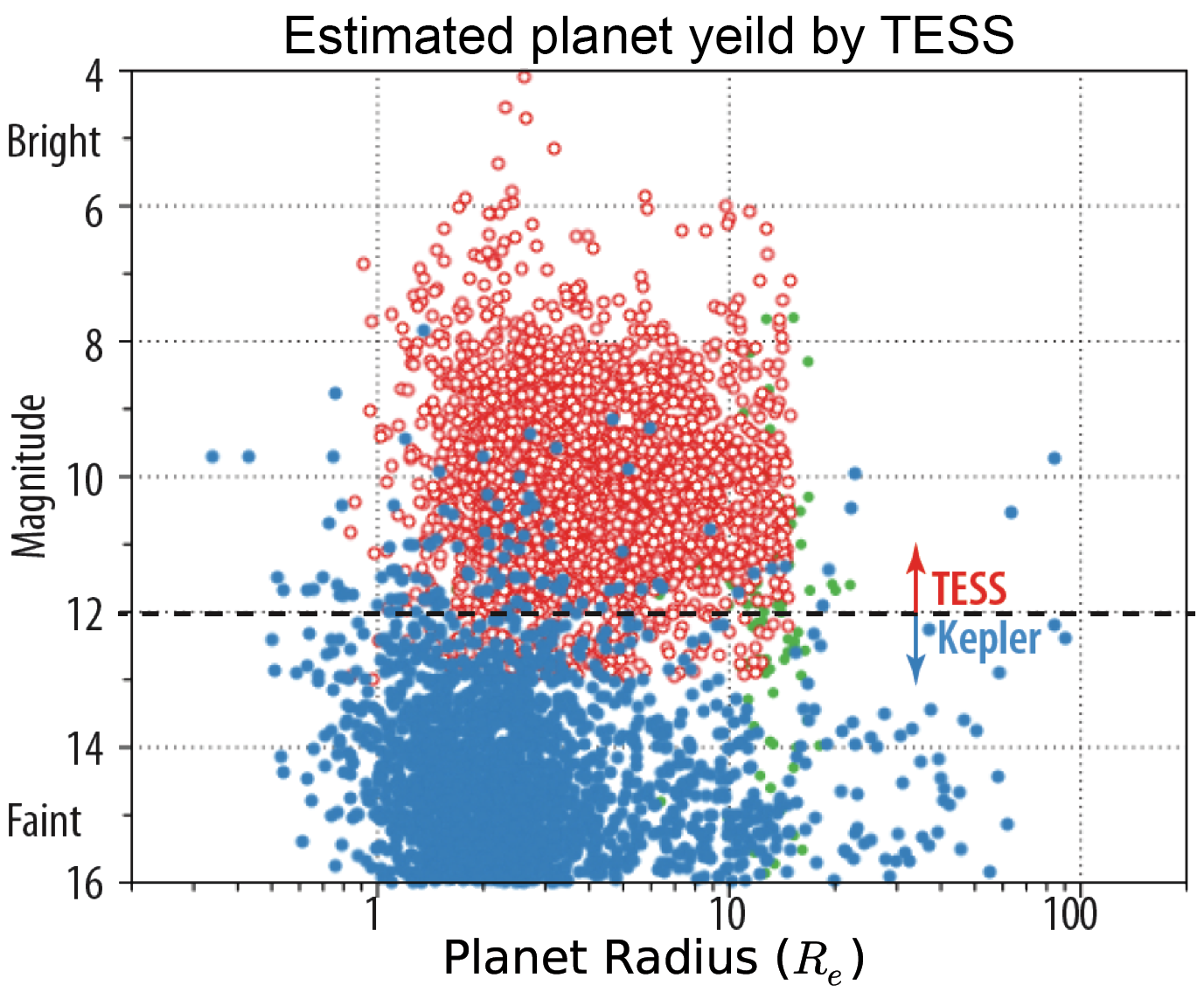}\\
    \end{tabular}
\label{fig:magv-r}
\end{figure}

\begin{figure}[p]
  \caption{Top Left: The predicted yield of planets from the Transiting Exoplanet Survey Satellite (TESS, figure published with permission from \citet{TESS2014}). The others show the current observable population (in the Chemical Census tier) with \echo{ }with stars brighter than J=10 and J=11. Note that differences in the \echo{ }and TESS plots are explained by differing catalogue versions, the TESS plot showing all targets (not just transiting) and our calculation of missing parameters (such as J magnitude) as described in section \ref{sec:Missing-info}}
  \centering
    \begin{tabular}{cc}
    \includegraphics[width=0.5\textwidth]{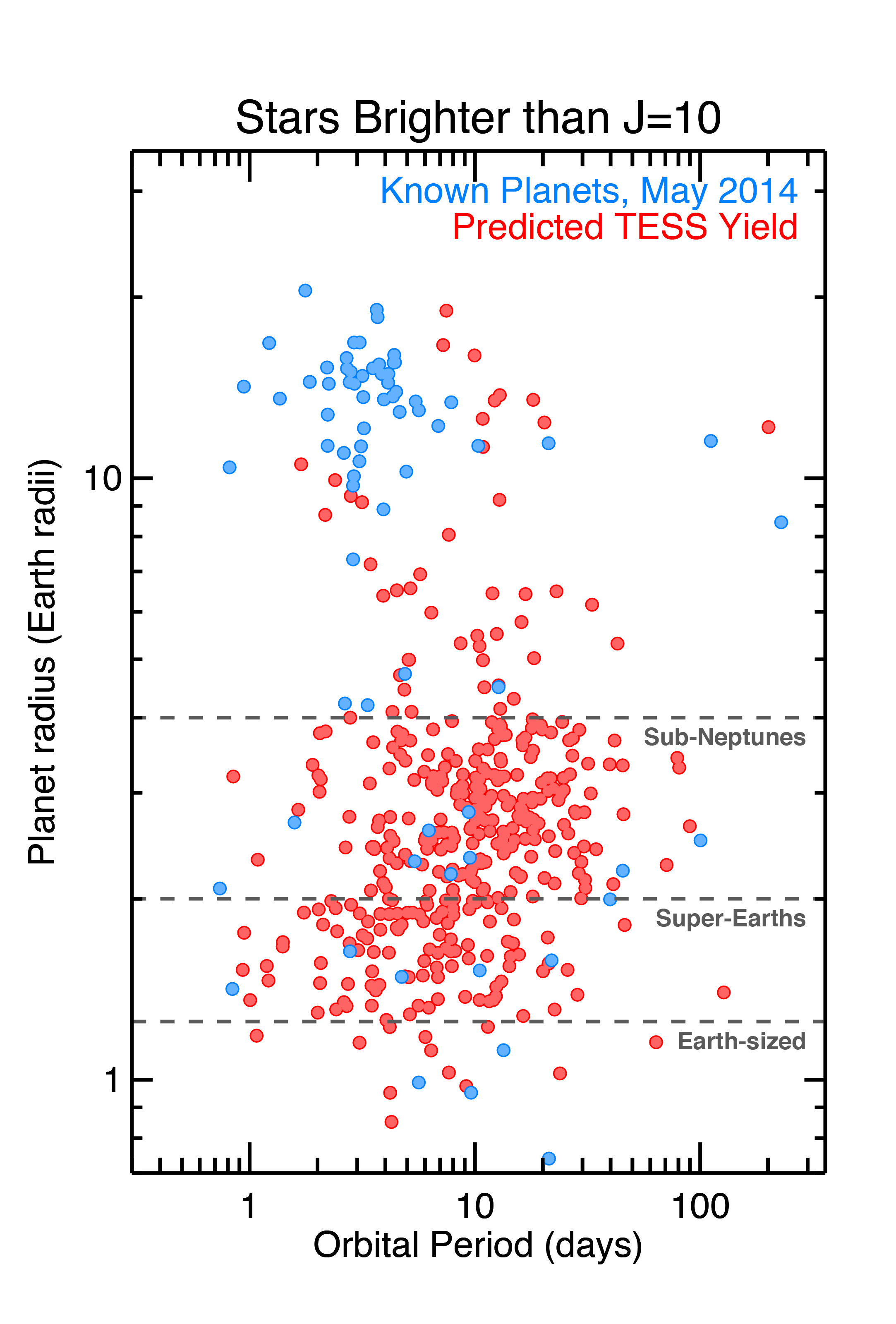}&
    \includegraphics[width=0.49\textwidth]{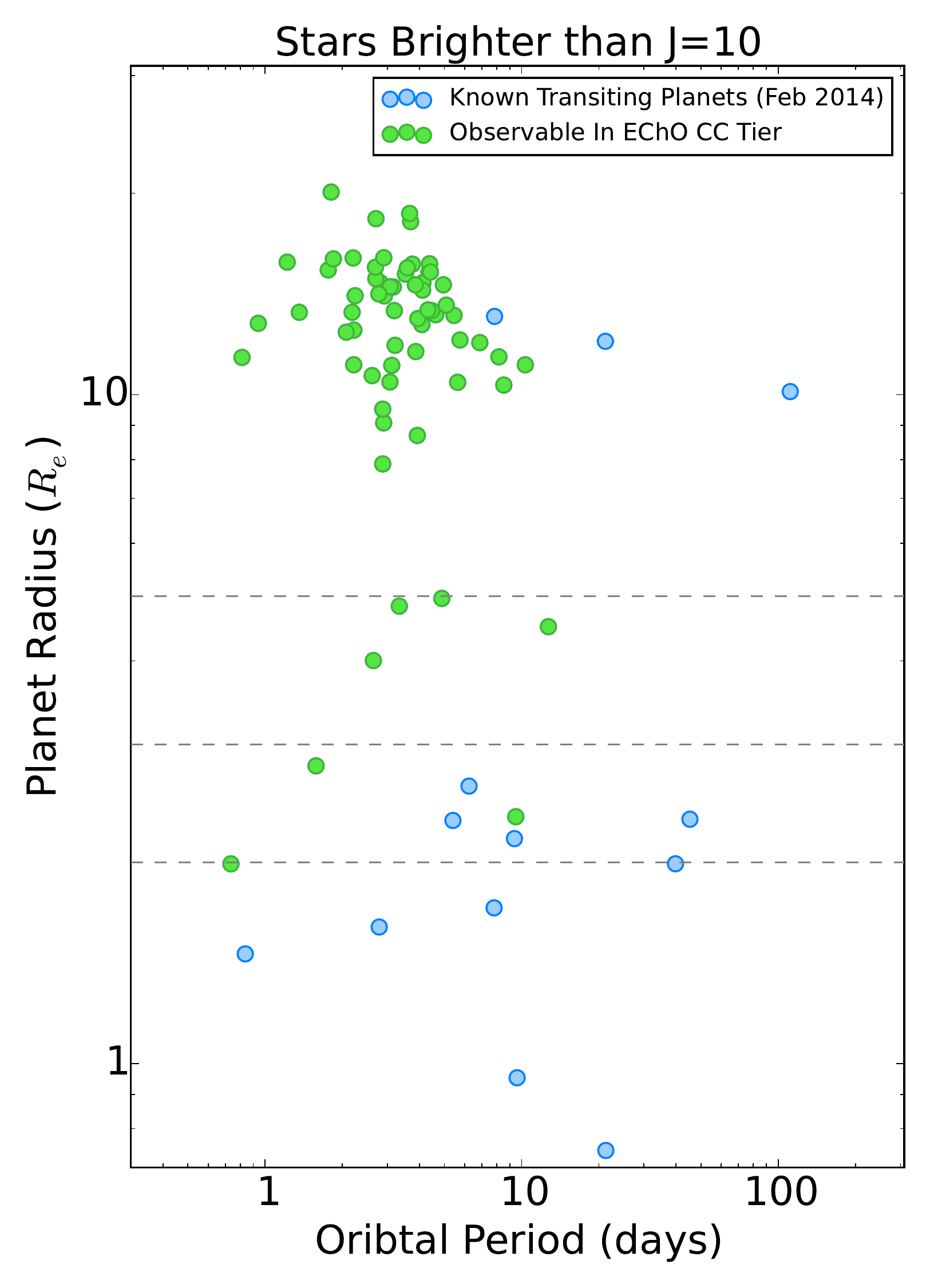}\\
    \includegraphics[width=0.49\textwidth]{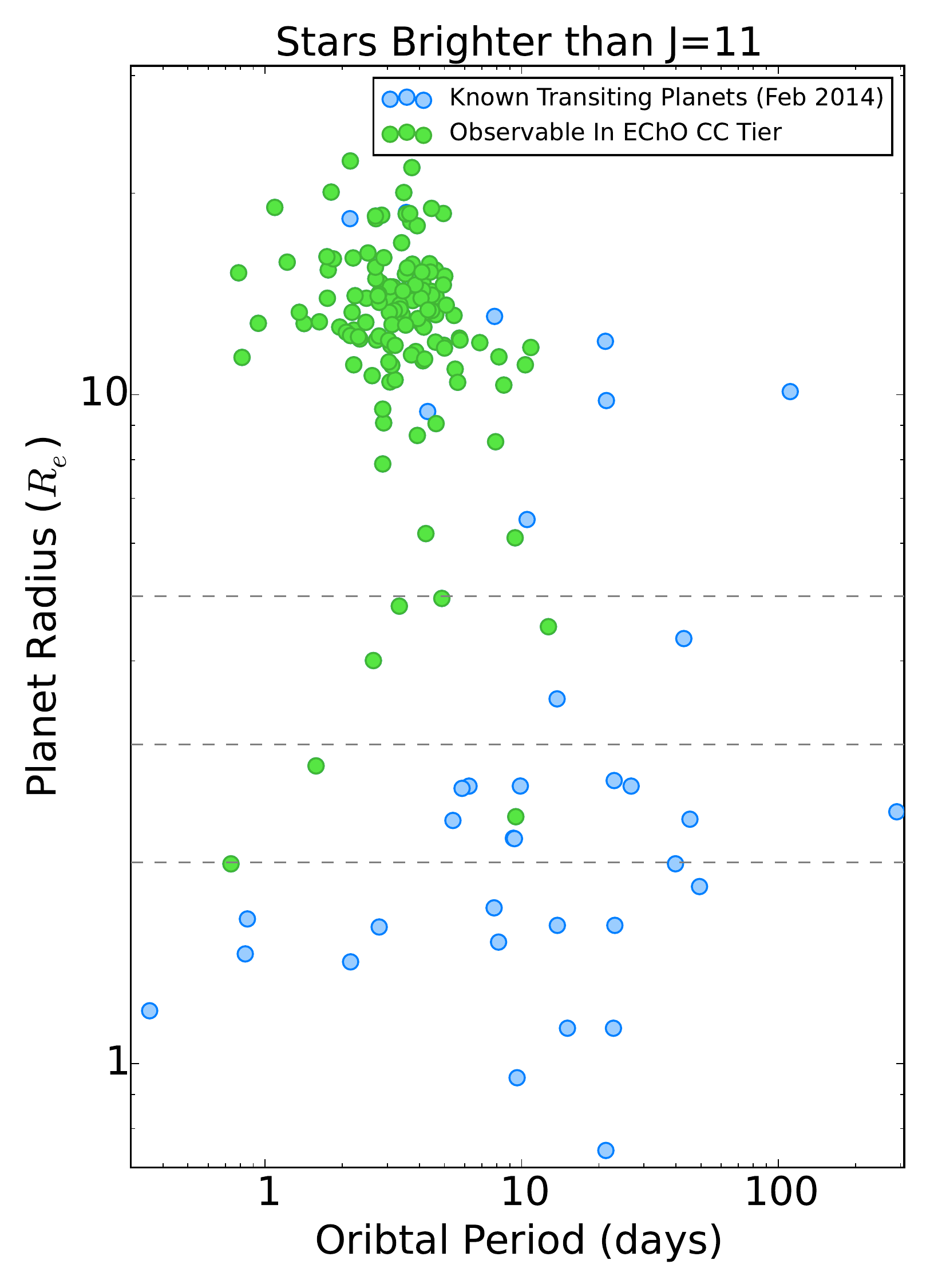}
    \end{tabular}
\label{fig:tess-period-radius}
\end{figure}

Fig. \ref{fig:magv-r} and \ref{fig:tess-period-radius} show the sample of todays targets that are observable with \echo{ } compared with the expected yield of the TESS mission \citep{TESS2014}; by stellar magnitude and planetary radius (Fig. \ref{fig:magv-r}) and orbital period and planetary radius (Fig. \ref{fig:tess-period-radius}).

\begin{figure}[p]
  \caption{Plot showing how the length of the mission affects the number of observable targets. The mission length required per target is calculated by taking the number of transits required meet the requirements of each tier times the period. Scheduling, overheads and total observation time are not taken into account (see discussion)}
  \centering
    \includegraphics[width=1\textwidth]{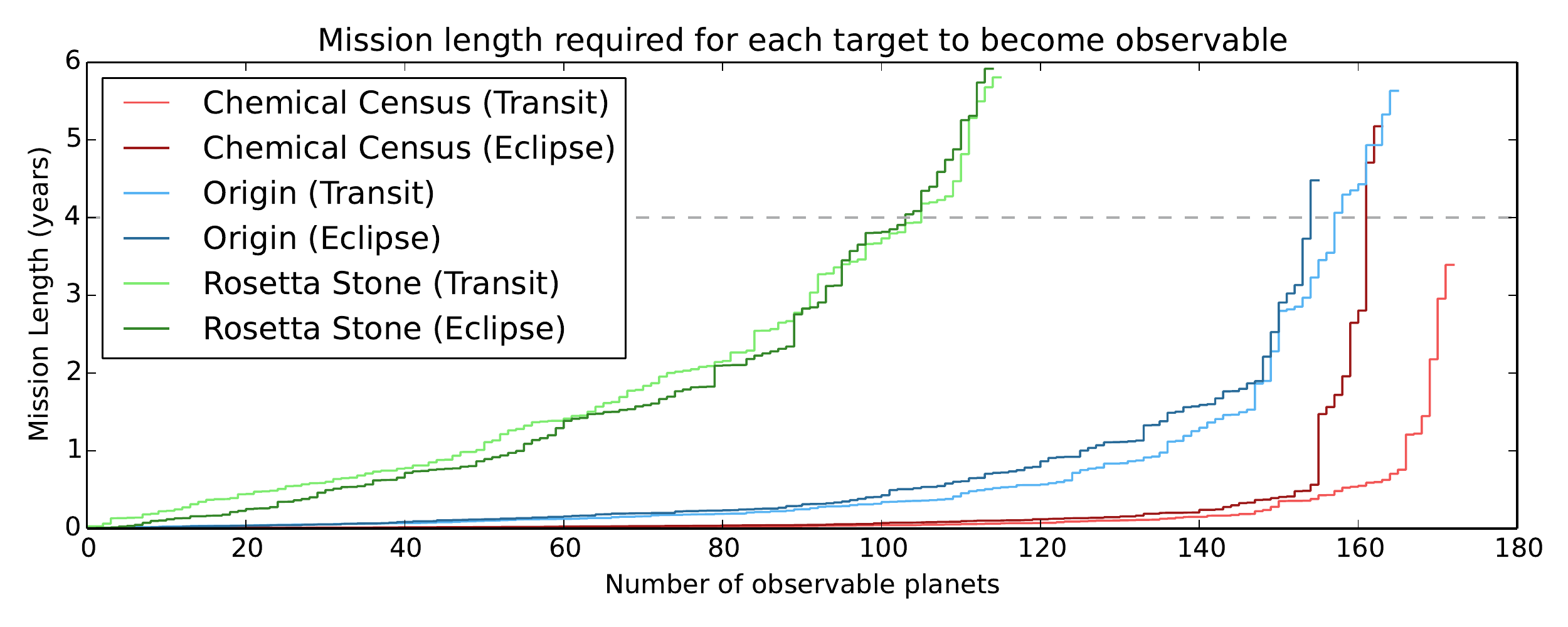}
\label{fig:mission-lifetime-plots}
\end{figure}

Fig. \ref{fig:mission-lifetime-plots} shows how the number of observable planets changes with mission lifetime (based on how long is needed for the required number of transit or eclipse events to occur). Overheads and scheduling are not considered here, see \citet{Morales14} and \citet{Garcia-Piquer14}. Note that the targets given here are the current sample which each observation tiers can handpick from based on the scientific benefit and scheduling.

\section{Discussion}
In this first iteration (V1.0) of the \echo{ }target list we show that a large number of characteristically diverse targets are observable today. 

As mentioned in our run conditions (\S \ref{sec:selection-conditions}), in this version we used a circular orbit assumption (transiting planets found so far normally have an eccentricity of less than 0.1). Eccentric targets have more complex orbital parameters (eg the planet temperature, transit duration, flux from the star and transit depth are all affected) which are not modelled automatically in the current versions of our simulators. Future software updates will incorporate the tools needed to simulate these planets; as more exoplanets are being discovered it is likely more targets will fall into this category increasing its importance. 

We do not consider scheduling or overheads in this exercise as we only generate the sample from which targets can be selected. Efficiency due to overheads is expected to be 80\% but scheduling conflicts and required observation time per target will be the major factors determining what can be observed (see \citet{Morales14} and \citet{Garcia-Piquer14}). The Rosetta Stone tier in particular requires a high number of transits for many of its targets which can be very demanding on the \echo{ }schedule.

In binary systems in which the planet orbits just one of the stars we can simulate these as normal, where the planet orbits the binary (ie the Kepler-47b system \citet{Orosz2012}) additional modelling will be required which will be included in a future version.

This work shows only the science achievable with today's (catalogue as of 25th February 2014) target sample. Survey missions from space like Gaia \citep{Perryman2013}, TESS \citep{TESS2014}, CHEOPS \citep{Broeg2013} and K2 \citep{K22014} and ground based surveys such as NGTS \citep{NGTS2013}, ESPRESSO \citep{ESPRESSO2010}, WASP \citep{WASP2006} and HAT \citep{HAT2004} are expected to yield thousands of transiting planets in the next five years. With many additional targets the choice of ideal candidates will be much greater in the future, adding to the efficiency of the \echo{ }mission. TESS in particular (launching 2017) is observing brighter targets than Kepler which are ideal candidates for \echo{ }(Fig. \ref{fig:magv-r}). See \citet{Micela14} for a detailed summary of each survey with respect to \echo{ }target selection.

\begin{figure}[htbp]
  \caption{The J-band limiting magnitudes for the different NIRSpec modes as a function of host star temperature. The lines represent the magnitude limit at which a source can be observed in the full wavelength range of the given mode. The dashed lines are for the high resolution gratings and solid lines for the medium resolution gratings. The plots are taken from the ESA JWST page on exoplanet transit spectroscopy with NIRSpec (http://www.cosmos.esa.int/web/jwst/exoplanets, retrieved 10th December 2014).}
  \centering
    \begin{tabular}{cc}
    \includegraphics[width=0.5\textwidth]{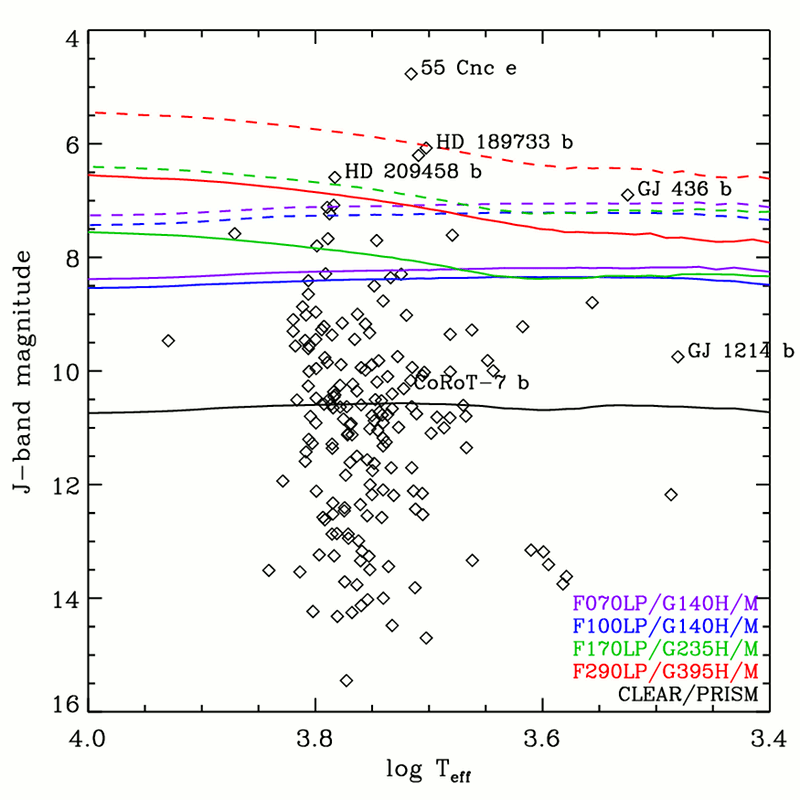}&
    \includegraphics[width=0.5\textwidth]{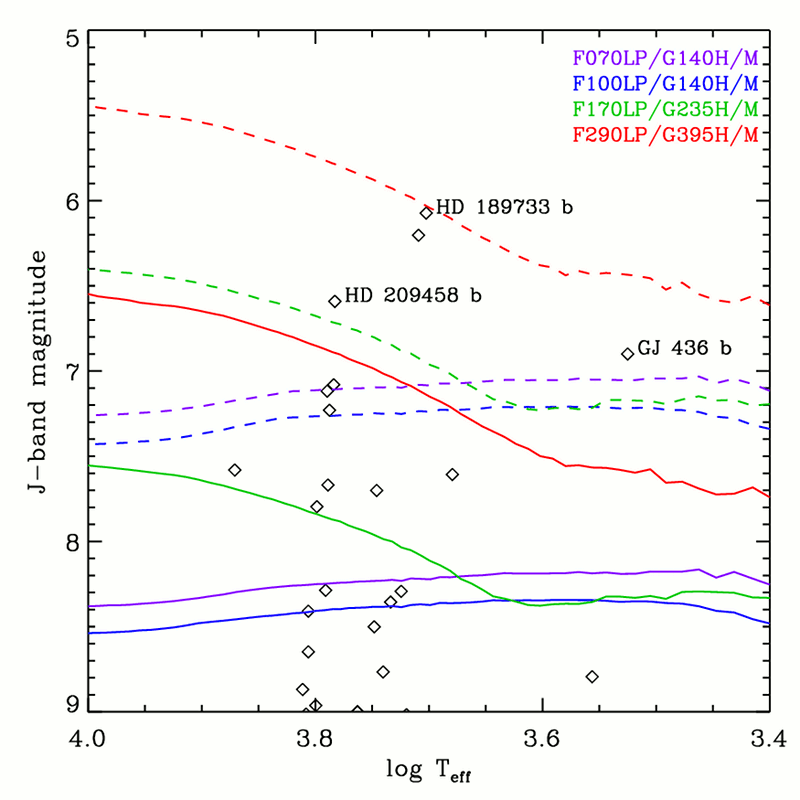}\\
    \end{tabular}
\label{fig:jwst-nirsepc}
\end{figure}

Both JWST (James Webb Space Telescope) and E-ELT (European Extremely Large Telescope) are expected on-line in the next decade and will take high resolution spectra of exoplanets. These are highly complementary and mutually beneficial to \echo{ }with JWST observing a few tens of planets at mid to high resolution and E-ELT providing ultra-high resolution spectroscopy of a few tens of planets in narrow bands. \echo{ }has been designed to look at the broader picture, surveying hundreds of exoplanets with instantaneous broad wavelength coverage. This instantaneous wavelength coverage allows \echo{ }observations to be corrected for stellar activity effects (like star-spots and faculae) which have a strong chromatic dependence \citep{Herrero2014, Micela14b, Danielski14, Scandariato2014}. Additionally some instrument on-board JWST and EChO are designed for different regimes e.g. JWST NIRSpec is particularly sensitive allowing high resolution spectroscopy in its spectral coverage but is therefore restricted to fainter targets (see Fig. \ref{fig:jwst-nirsepc}). Whilst \echo{ }is optimised for brighter stars around K $<$ 9 \citep{EChOYB} JWST NIRSpec is optimised for targets roughly K $>$ 8.5 magnitude.

\section{Conclusion}
EChO has been designed as a 1 metre dedicated survey mission for transit and eclipse spectroscopy capable of observing a large, diverse and well-defined planet sample within its four year mission lifetime, our results show that the majority of this diversity can be achieved with today's target sample. 

173 of today's targets can be observed in \echo's broadest survey tier (Chemical Census, R = 50 at $\lambda < 5 \mu m$, SNR = 5) in transit and/or eclipse, 162 are observable in both. This sample covers a wide range of planetary and stellar sizes, temperatures, metallicities and semi-major axes. This excludes the recent discovery of over 700 planets \citep{2014Lissauer, 2014Rowe} that will be added in a future version.

Out of these 173, the majority (165) can be observed in transit or eclipse (148 in both) at the higher spectral resolving power and SNR of the Origin tier (R = 100 at $\lambda < 5 \mu m$, SNR = 10). Dedicated studies show that an accurate retrieval can be performed out of origin targets, so that the physical causes of said diversity can be identified (\citet[][in prep]{Tessenyi2014}, \citet{Barstow2014}).

For a subset of these, we can push the spectral resolving power to R = 300 at $\lambda < 5 \mu m$ at SNR = 20 so that a very detailed knowledge of the planets can be achieved (Rosetta Stone). Said knowledge will include spatial and temporal resolution enabling studies of weather and climate, as well as very refined chemical composition of these atmospheres to penetrate the intricacies of equilibrium and non-equilibrium chemistry and formation. While today there are 132 targets capable of being observed in this tier in transit or eclipse (of which 78 can be done in both), as the Rosetta Stone tier is very demanding of the \echo{ }schedule, this is the large sample from which the target 10-20 planets can be chosen from.

\echo's unique contribution to exoplanetary science is in identifying the main constituents of hundreds of exoplanets in various mass/temperature regimes, meaning that we will be looking no longer at individual cases but at populations of planets. Such a universal view is critical if we truly want to understand the processes of planet formation and evolution and how they behave in various environments.

\begin{acknowledgements}
R. Varley is Funded by a UCL IMPACT Studentship, I. Waldmann is funded by the UK Space Agency and STFC, J. C. Morales is funded by a CNES fellowship. G. Tinetti is a Royal Society URF, G. Micela acknowledges support by the ASI/INAF contract I/022/12/0.

We would like to thank Vincent Coudé du Foresto  and Jean-Philippe Beaulieu for their help with this work.
\end{acknowledgements}

\bibliographystyle{aa}
\bibliography{echo-targetlist-paper}   

\end{document}